\documentclass[prd,onecolumn,showpacs,nofootinbib,fleqn]{revtex4}

\usepackage[utf8]{inputenc}
\usepackage{natbib}
\usepackage{amsmath}
\usepackage{amssymb}
\usepackage{amsfonts}
\usepackage{xfrac}
\usepackage{graphicx}
\usepackage{subfigure}
\usepackage{hyperref}
\usepackage{todonotes}
\usepackage[normalem]{ulem}

\newcommand{\diff}{\mathrm{d}}
\newcommand{\defeq}{:=}
\renewcommand{\vector}[1]{\boldsymbol #1}
\newcommand{\Mpl}{M_\mathrm{Pl}}
\newcommand{\order}[1]{\mathcal{O}\left(#1\right)}
\DeclareMathOperator{\Beta}{\mathrm{Beta}}
\newcommand{\stn}{\frac{\mathrm{S}}{\mathrm{N}}}
\newcommand{\three}[3]{$[#1\,(#2)\,#3]$}
\newcommand{\curv}{\mathcal{R}}
\newcommand{\cs}{c_\mathrm{s}}
\newcommand{\umax}{|u|_\mathrm{max}}
\newcommand{\smax}{|s|_\mathrm{max}}
\newcommand{\fnl}{f_{\scriptscriptstyle \mathrm{NL}}}
\newcommand{\dpp}{\frac{\Delta \mathcal{P}_\curv}{\mathcal{P}_\curv}}
\newcommand{\sdpp}{\Delta \mathcal{P}_\curv/\mathcal{P}_\curv}
\newcommand{\dppt}{\frac{1}{\umax}\dpp}
\newcommand{\perms}{\stackrel{\mathrm{perms}}{\ldots}}
\newcommand{\cperms}{\stackrel{\mathrm{cyclic}}{\ldots}}

\begin{document}

\title{Robust predictions for an oscillatory bispectrum in Planck 2015 data from transient reductions in the speed of sound of the inflaton}

\author{Jesús Torrado$^{1,2}$, Bin Hu$^{3,4}$ and Ana Achúcarro$^{5,6}$}
\affiliation{
\smallskip
$^{1}$ Astronomy Centre, University of Sussex, Brighton BN1 9QH, UK \emph{(current address)}\\
$^{2}$ Institute for Astronomy, University of Edinburgh, Blackford Hill, Edinburgh EH9 3HJ, UK\\
$^{3}$ Department of Astronomy, Beijing Normal University, Beijing, 100875, China\\
$^{4}$ Institut de Ciències del Cosmos (ICCUB), Universitat de Barcelona (IEEC-UB), Martí i Franquès 1, E08028 Barcelona, Spain\\
$^{5}$ Institute Lorentz, Leiden University, PO Box 9506, Leiden 2300 RA, The Netherlands\\
$^{6}$ Dept.\ of Theoretical Physics, University of the Basque Country UPV-EHU, 48080 Bilbao (Spain)}

\begin{abstract}
We update the search for features in the Cosmic Microwave Background (CMB) power spectrum due to transient reductions in the speed of sound, using Planck 2015 CMB temperature and polarisation data. We enlarge the parameter space to much higher oscillatory frequencies of the feature, and define a robust prior independent of the ansatz for the reduction, guaranteed to reproduce the assumptions of the theoretical model. This prior exhausts the regime in which features coming from a Gaussian reduction are easily distinguishable from the baseline cosmology. We find a fit to the $\ell\approx20$--$40$ minus/plus structure in Planck TT power spectrum, as well as features spanning along higher $\ell$'s ($\ell\approx100$--$1500$).
None of those fits is statistically significant, either in terms of their improvement of the likelihood or in terms of the Bayes ratio. For the higher-$\ell$ ones, their oscillatory frequency (and their amplitude to a lesser extent) is tightly constrained, so they can be considered robust, falsifiable predictions for their correlated features in the CMB bispectrum. We compute said correlated features, and assess their signal-to-noise and correlation to the ISW-lensing secondary bispectrum. We compare our findings to the shape-agnostic oscillatory template tested in Planck 2015, and we comment on some tantalising coincidences with some of the traits described in Planck's 2015 bispectrum data.
\end{abstract}


\maketitle

\section{Introduction}

The Planck collaboration \cite{Adam:2015rua} has released all the data taken by the survey, including polarisation power spectrum and some results of the analysis of the bispectrum. However, a likelihood for the CMB bispectrum has not been released for public use. The analyses carried out by the Planck collaboration in the context of primordial fluctuations have not found any strong deviation from the predictions of the canonical single-field slow-roll inflation paradigm. In particular, they found no significant deviation from the vanilla power-law power spectrum \cite{Ade:2015lrj}, neither a detection of any shape of primordial non-Gaussianity \cite{Ade:2015ava}. Some hints are reported for small deviations on both data sets, but always in the low signal-to-noise regime, under the significance necessary to claim a detection. Some of those hints persist from Planck 2013 \cite{Planck:2013jfk, Ade:2013ydc} through Planck 2015 (and even WMAP \cite{2012JCAP...12..032F}), such as a dip at $\ell\approx 20$ and some small features in the CMB temperature bispectrum, that have been deemed interesting by the Planck collaboration.

Many of the extensions of canonical single-field slow-roll inflation predict \cite{Chen:2006xjb} correlated features in both the 2- and 3-point correlation functions.\footnote{There is very extensive literature on this subject; we refer the reader to the recent review \cite{Chluba:2015bqa}.} Notably, in a few cases, the correlations can be computed explicitly \cite{Achucarro:2012fd,Gong:2014spa,Palma:2014hra,Chluba:2015bqa,Mooij:2015cxa}.
When models with correlated features are tested against the data in a joint approach for different observables at the same time, the significance of possible fits is expected to increase, as has been reported in particular for oscillatory feature searches combining CMB power spectrum and bispectrum  \cite{Fergusson:2014hya,Fergusson:2014tza,Meerburg:2015owa} (see also \cite{Appleby:2015bpw} for a model-independent approach).

This motivates us to update our ongoing search \cite{Achucarro:2013cva,Achucarro:2014msa,Hu:2014hra} for features produced by transient reductions in the speed of sound of the inflaton \cite{Achucarro:2012fd} with the new Planck 2015 temperature and polarisation power spectrum data, in preparation for a joint search including bispectrum data. As a part of it, we have re-evaluated the prior of our search to ensure theoretical self-consistency in a more efficient way (imposed a priori, not a posteriori) and enlarged the parameter space such that it covers all the configurations for which the feature is \emph{distinguishable} from the baseline cosmology. With the results of this updated search, we formulate predictions for the CMB bispectrum that are \emph{robust}, i.e.\ are guaranteed to be theoretically self-consistent and have a very narrow range of oscillatory frequencies. They are also fundamentally different to the oscillatory bispectrum templates tested by Planck so far, in that the oscillations in the squeezed limit are out of phase by $\pi/2$ with those on the equilateral and folded limits.

The present paper is structured as follows: we begin by reviewing the theoretical framework for this family of inflationary features, and describe their shape in the CMB observables (section \ref{sec:theoryreview}); then, we present our ansatz for the speed of sound reduction (section \ref{sec:gaussian}) and discuss the prior that we will employ in our sampling (section \ref{sec:prior}). After discussing the data sets and methodology with which our search has been conducted (section \ref{sec:methodology}), we present and discuss our results for the CMB power spectrum (section \ref{sec:powerspectrum}), and draw from them predictions for the CMB bispectrum (section \ref{sec:bispectrum}) which are discussed in the context of Planck's search for non-Gaussianity. Finally, we discuss the relevance of our findings and prospects for searches for features of this kind (section \ref{sec:conclusion}). The numerical tools used to carry out CMB bispectrum computation and forecasts are described in appendix \ref{app:bispectrum}.

\section{Theoretical model and prior}\label{sec:theory}

\subsection{Review of the theoretical model}
\label{sec:theoryreview}

We work in the framework of effective field theory of inflationary perturbations \cite{Cheung:2007st}, 
described in terms of the Goldstone boson of time diffeomorphisms, $\pi(t,\vector{x})$. 
This is related to the adiabatic curvature perturbation linearly: $\curv(t,\vector{x})=-H(t)\pi(t,\vector{x})$, with $H\defeq {\dot{a}}/{a}$ and $a$ is the scale factor (from now on, we use natural units, $\hbar=c=1$, define the reduced Planck mass as $\Mpl^{-2}\defeq8\pi G$, and denote physical time derivatives with an overdot, $\dot {\phantom{x}} \defeq \diff/\diff t$).

The effective quadratic action for $\pi$ reads
\begin{equation}
\label{eq:S2}
S_2 = \Mpl^2 \int \diff^4x\,\frac{\epsilon\, a^3 H^2}{\cs^2}
      \left\{\dot{\pi}^2 - \cs^2 \frac{(\nabla \pi)^2}{a^2}\right\}
\,,
\end{equation}
where $\epsilon\defeq -{\dot{H}}/{H^2}$ and the time-dependent speed of sound $\cs$ that appears in the action accounts for the effect of the heavy components of the field space that are made implicit by the effective field theory.

In order to get a physical grasp of the significance of a speed of sound reduction, carrying out explicitly the integration of the heavy mode in a 2-field scenario, one gets \cite{Achucarro:2010jv}
\begin{equation}
c_s = \left(1+\frac{4\dot{\theta}^2}{M^2 - \dot{\theta}^2}\right)^{-2}
\,,
\end{equation}
where $\dot{\theta}$ is the angular velocity 
of the background trajectory along the approximate minimum of the potential, and
$M^2$ would be the mass squared 
of the heavy modes perpendicular to that trajectory if the trajectory were straight. Thus, soft, adiabatic \emph{turns} in the inflationary trajectory in field space result in transient reductions of the speed of sound.\footnote{Sufficiently sharp turns would violate the adiabatic condition that prevents quanta of the heavy degrees of freedom from being produced \cite{Cespedes:2012hu,Achucarro:2012yr}: $|\dot{c}_\mathrm{s}|\ll M|1-\cs^2|$.}

We can rewrite the quadratic action \eqref{eq:S2} as
\begin{equation}
S_2 = S_{2,\mathrm{free}} + \Mpl^2 \int \diff^4x\,\epsilon\, a^3 H^2 (-u \dot{\pi}^2)
\,,    
\end{equation}
where $S_{2,\mathrm{free}}\defeq S_2(\cs=1)$ and we have re-parametrised the varying speed of sound as \cite{Seery:2005wm}
\begin{equation}
  u\defeq 1-\frac{1}{\cs^2}
\,,
\end{equation}
which departs from zero towards \emph{negative} values when the speed of sound departs from unity. Treating the transient speed of sound as a small perturbation of the free action and using the \emph{in-in} formalism \cite{Weinberg:2005vy}, one sees that mild changes in the speed of sound seed \emph{features} in the primordial power spectrum of curvature perturbations as \cite{Achucarro:2012fd}
\begin{equation}
\label{eq:dpp}
\dpp(k) = k \int_{-\infty}^0 \diff\tau\,u(\tau)\sin(2k\tau)
\,,
\end{equation}
where $\mathcal{P}_\curv={H^2}/({8\pi^2\epsilon\Mpl^2})$ is the featureless nearly-scale-invariant power spectrum corresponding to the constant case $\cs=1$ ($u=0$), and where $u(\tau)$ departs briefly and softly from zero and back, and $\tau$ is the conformal time.

One can also write the cubic action for the adiabatic mode:
\begin{equation}
S_3 = \Mpl^2 \int \diff^4x\,\epsilon\, a^3 H^2
      \left\{-2\,(1-u)\,s\,H\pi\dot{\pi}^2 -
             u\,\dot{\pi}\left[\dot{\pi}^2-\frac{(\nabla \pi)^2}{a^2}\right]\right\}
\,,
\end{equation}
where we have introduced the relative derivative of the speed of sound
\begin{equation}
  s\defeq \frac{1}{H}\frac{\dot{c}_\mathrm{s}}{\cs}
\,.
\end{equation}
In the cubic action above, two important assumptions have been made:
\begin{itemize}
\item Slow-roll contributions still present in constant-speed-of-sound scenarios are neglected. They come at order $\order{\epsilon^2,\eta^2}$ \cite{Maldacena:2002vr}, so in order for this assumption to be correct (i.e.\ this here being the main contribution to the cubic action), \emph{at least one of} $u$ or $s$ must be significantly larger than the slow-roll parameters, at least at their maximum deviation from zero.
\item The cubic action due to the speed of sound reduction is treated perturbatively, so if we want to be sure that higher order terms can be neglected, \emph{both} the speed of sound and its change rate as they appear in the cubic action, $u$ and $s$, must be significantly smaller than $1$ at their maxima.
\end{itemize}
Summarising:
\begin{equation}
  \label{eq:uslimits}
  \max\left(\epsilon,|\eta|\right) \ll \max\left(\umax,\smax\right) \ll 1
\,.
\end{equation}
In section \ref{sec:prior}, we discuss how to impose those bounds in a natural way.

It is easy to check that the perturbative limit on $\smax$ ensures that the consistency conditions derived in \cite{Cespedes:2012hu,Adshead:2014sga,Cannone:2014qna} are comfortably satisfied, setting a limit to the sharpness of the reduction at least as stringent as the ones found in those references. Thus, as long as our prior duly imposes those bounds in $\umax$ and $\smax$, we eliminate the risk of fitting to the data features whose computation can be found a posteriori not to be theoretically consistent.\footnote{This is different from the treatment in our previous work \cite{Achucarro:2014msa,Achucarro:2013cva}, and also in \cite{Miranda:2013wxa,Planck:2013jfk} for steps in the potential.}

From the cubic action above, again using the \emph{in-in} formalism, one can compute the main contribution to the bispectrum of the curvature perturbations \cite{Achucarro:2012fd}
\begin{equation}
  \label{eq:bispSRFT}
  B_\curv(k_1, k_2, k_3) = \frac{(2\pi)^4\mathcal{A}_s^2\Mpl^6}{(k_1k_2k_3)^2}
  \sum_{i=0}^2 c_i(k_1, k_2, k_3) \left(\frac{k_t}{2}\right)^i
  \left(\frac{\diff}{\diff k_t/2}\right)^i\dpp(k_t/2)
  \,,
\end{equation}
where $k_t \defeq k_1+k_2+k_3$. The scale-independent \emph{shape coefficients} $c_i$ are:
\begin{subequations}
\label{eq:bispeccoeffs}
\begin{align}
  c_0 &\defeq -\frac{1}{k_t^2}\left(k_1k_2+\cperms\right)+
              \frac{ 1}{4}\frac{1}{k_t}\left(\frac{k_1^3}{k_2k_3}+\cperms\right)
              -\frac{3}{2}\frac{1}{k_t}\left(\frac{k_1k_2}{k_3}+\cperms\right)+
              \frac{1}{4}k_t\left(\frac{1}{k_1}+\cperms\right)
              -\frac{5}{4}\,,\\
  c_1 &\defeq 
         \frac{1}{k_t^2}\left(k_1k_2+\cperms\right)-\frac{19}{32}
        + \underbrace{
              \frac{19}{32} - \frac{1}{4}\frac{1}{k_t}\left(\frac{k_2^2+k_3^2}{k_1}+\cperms\right)
              }_{c_{1,\mathrm{sq}}}\,,\\
  c_2 &\defeq \frac{1}{4}\frac{1}{k_t^2}(k_1^2 + k_2^2 +k_3^2)\,,
\end{align}
\end{subequations}
where $\cperms$ means the 2 remaining cyclic permutations of the $k_i$ (the missing $k_i$'s in a term are understood to be implicit, e.g.\ $k_1k_2+\cperms\defeq k_1k_2+k_2k_3+k_3k_1$).

Notice that, unlike in most of the literature, we are not extracting an overall amplitude $\fnl$ in front of the bispectrum.  We could use $\umax$ as a proxy for $\fnl$, redefining $(\sdpp)^\star \defeq \sfrac{1}{\umax}\sdpp$. Also, we notice that there is a non-separable prior on $\umax$ and $\smax$ determined by eq.\ \eqref{eq:uslimits} (and developed in section \ref{sec:prior}). This non-separability of the amplitude from the rest of the shape parameters should be taken into account when fitting this template to the data, since the range of amplitudes $\umax$ allowed by the prior depends on the value of $\smax$ of the tested template (see section \ref{sec:prior}).

The main result from \cite{Achucarro:2012fd} is thus that \emph{features in the power spectrum and the bispectrum are correlated in a very simple, analytic way}, and that both are easily expressed in terms of a mild, transient reduction of the speed of sound $u(\tau)$ of the adiabatic mode. It is worth remarking that both observables were re-computed in the same theoretical framework using the generalised slow-roll formalism in \cite{Achucarro:2014msa}, and they were found to be consistent with the expressions above, with agreement improving as the reductions get sharper (large $\smax$), i.e.\ the regime where the generalised slow-roll approximation works best.

Let us discuss a little the appearance of those features in both observables. Let as assume that the speed of sound reduction happens around a particular instant $\tau_0$, that we will define as the instant of maximum reduction: $u(\tau_0)\defeq-\umax$. The rate of change $s$ being limited from below by the slow-roll parameters means that the reduction must be approximately localised around $\tau_0$. The Fourier transform in eq.\ \eqref{eq:dpp} turns that localisation into a linear-in-$k$ oscillatory factor $\sin(2k\tau_0)$ for the power spectrum feature, with possibly a small phase if the reduction is not symmetric around $\tau_0$. The finite span in $\tau$ of the reduction imposes a finite envelope on top of those oscillations, the details of which (weight of the tails, symmetry) are determined by the particular shape of $u(\tau)$.

In the bispectrum, all this remains true, the oscillatory factor being $\sin(k_t\tau_0)$. The variation along total scale $k_t\defeq k_1+k_2+k_3$ is given mainly by $\sdpp$ and its derivatives, so when observed along $k_t$ in a particular direction (i.e.\ a particular triangular configuration), the feature will look similar to that on the power spectrum: an enveloped oscillation. The amplitude and phase do change across different triangular configurations: the \emph{central} configurations (i.e.\ those away from the squeezed limit, including the \emph{equilateral} and \emph{folded} limits) are dominated by the term with the second derivative, and may receive additional contributions from the rest of the terms (mostly from the zeroth derivative) if the reduction is not specially sharp, i.e.\ $\smax\sim\umax$. The \emph{squeezed limit} is completely defined by the term $c_{1,\mathrm{sq}}$ in the first derivative alone, which diverges towards that limit as the inverse of the smallest wave number. Despite there apparently being squeezed contributions from $c_0\cdot\sdpp$, they cancel out, in agreement with the consistency condition \cite{Maldacena:2002vr,Creminelli:2004yq}.

\begin{figure}[ht]
\centering
\subfigure[
\label{fig:primordialP}
]{\includegraphics[width=0.505\textwidth]{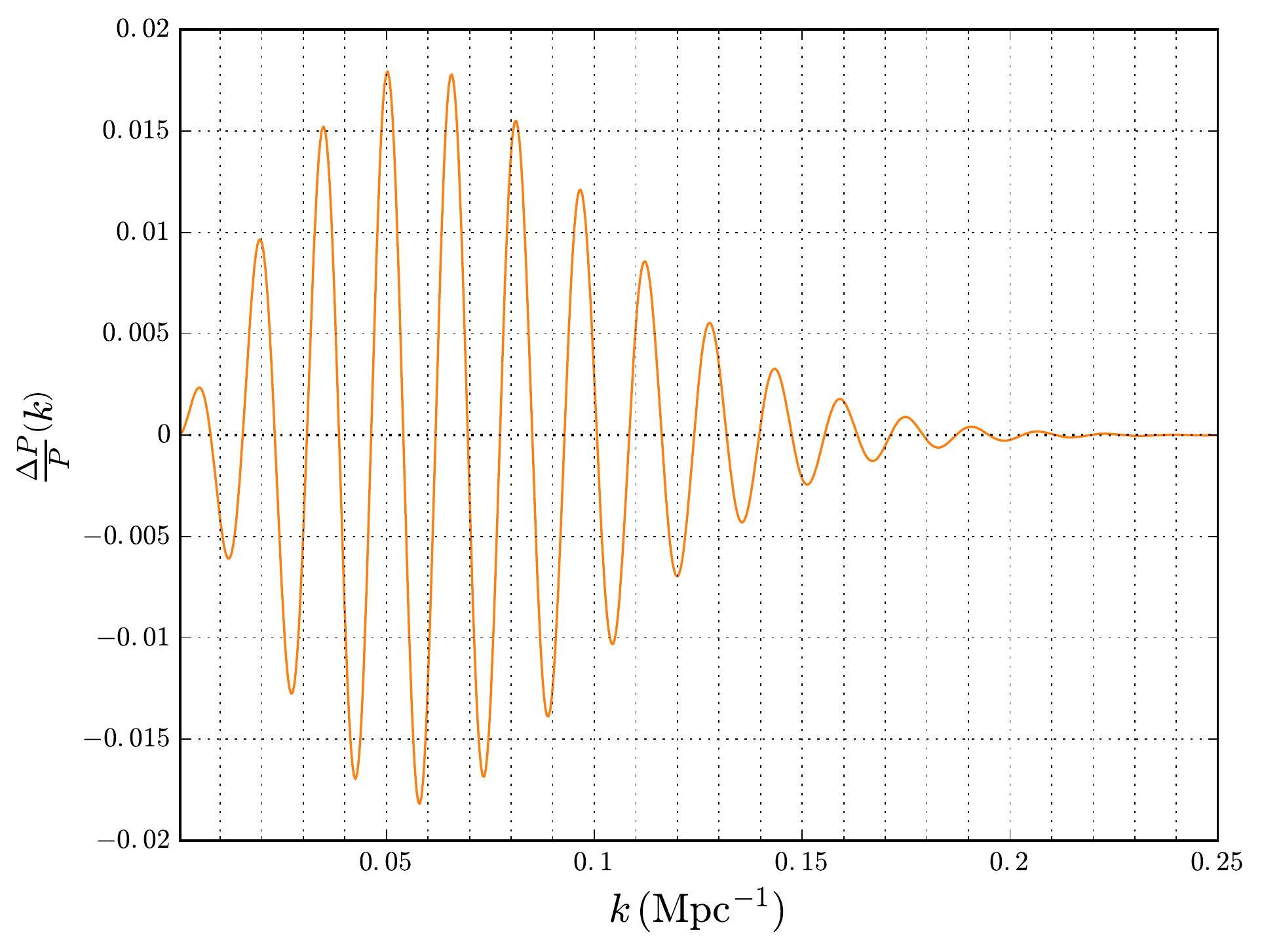}}
\subfigure[
\label{fig:primordialB}
]{\includegraphics[width=0.48\textwidth]{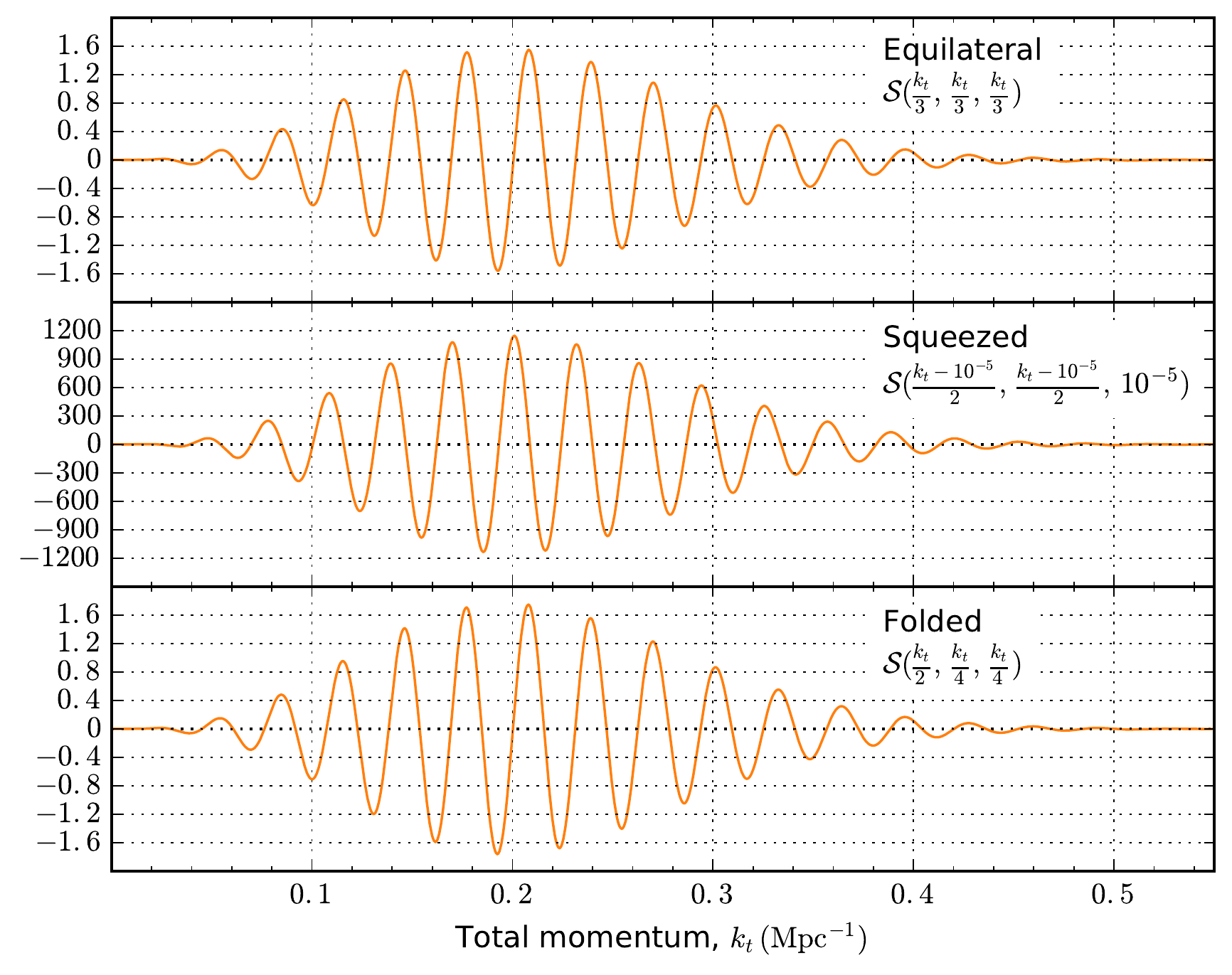}}
\caption{
\label{fig:primordial}
Features in the primordial power spectrum \ref{fig:primordialP} and bispectrum \ref{fig:primordialB} (with ${(2\pi)^4\mathcal{A}_s^2\Mpl^6}(k_1k_2k_3)^{-2}\cdot\mathcal{S}\defeq\mathcal{B}_\curv$) from a Gaussian reduction in the speed of sound, eq.\ \eqref{eq:gaussN}, with parameters $B=-0.024$, $\log\beta=5.6$ and $\tau_0=-203$, corresponding to one of our \textit{maxima a posteriori} (see table \ref{tab:maps}). Notice the linear oscillation along the (total) scale for the (bi)spectrum, and the ${\pi}/{2}$ phase difference between the squeezed and both the equilateral and folded shapes of the bispectrum, as discussed at the end of section \ref{sec:theory}.
}
\end{figure}

Due to the order of the derivatives, \emph{the oscillations in the squeezed limits are out of phase by ${\pi}/{2}$ with respect to those at the central configurations}. This can be seen clearly in figure \ref{fig:primordialB}, comparing the middle plot with the upper and lower ones. This is the main difference with the shapes tested so far on the Planck bispectrum in the 2013 \cite{Ade:2013ydc} and 2015 \cite{Ade:2015ava} data releases (see section \ref{sec:bispectrum}), which are all proportional to $\sin(\omega k_t+\phi)$, where the phase is the same for all triangular configurations.

All the statements above about the characteristics of the features are independent from the particular ansatz chosen for the reduction, and are illustrated in figure \ref{fig:primordial}. In the next section, we present our case study: a Gaussian reduction of the speed of sound.\footnote{As an alternative approach, one could parametrise the equation of state of the inflaton and derive from it the variation in the speed of sound, as in \cite{Gariazzo:2016blm}.}

\subsection{Gaussian ansatz for the reduction}
\label{sec:gaussian}

As in our previous work \cite{Achucarro:2013cva,Achucarro:2014msa,Hu:2014hra}, we propose a reduction in the speed of sound as a Gaussian in $\mathrm{e}$-folds (or, equivalently, in physical time):\footnote{Choosing a Gaussian in $\tau$ would have been problematic: it would never be exactly zero by $\tau_0$, as required for the expression of the bispectrum in eq.\ \eqref{eq:bispSRFT}.}
\begin{equation}
  \label{eq:gaussN}
  u(\tau) \defeq B \exp\left\{-\beta(N-N_0)^2\right\}
          = B \exp\left\{-\beta\left(\log\frac{\tau}{\tau_0}\right)^2\right\}
\,.
\end{equation}
This reduction is parametrised by its maximum intensity $B<0$, a sharpness $\beta>0$ and an instant of maximum reduction $\tau_0<0$ (or, equivalently, $N_0$). As explained in the last section, $\tau_0$ is the instant around which the reduction is localised. The intensity and the sharpness here are related to the maxima in the reduction $\umax$ and its rate of change $\smax$ as
\begin{equation}
  \label{eq:paramsrelations}
  \umax = -B
  \qquad\text{and}\qquad
  \smax = \sqrt{\frac{\beta}{2}} \frac{-B}{e^\frac{1}{2}-B}
\,.
\end{equation}

Notice that this functional form has naturally 3 parameters only, $(B,\beta,\tau_0)$, exactly as many as we used in the last section to characterise a reduction in a model independent way: $(\umax,\smax,\tau_0)$. Also, a Gaussian is one of the simplest functions that softly departs from zero and returns.

\subsection{Prior}
\label{sec:prior}

In our previous work \cite{Achucarro:2013cva,Achucarro:2014msa,Hu:2014hra}, we imposed a uniform prior directly on the parameters of the Gaussian reduction, and checked that $\smax\ll 1$ a posteriori. Since $\smax$ depends on both $\beta$ and $B$ simultaneously, see eq.\ \eqref{eq:paramsrelations}, a rectangular region in $(B,\beta)$ does not map nicely into one in $(\umax,\smax)$, where the prior motivated by eq.\ \eqref{eq:uslimits} should be imposed. Thus, in those papers we successfully explored the parameter region of interest, but in an inefficient manner: regions of the parameter space not allowed by the theory were thoroughly explored to later be thrown away.

\begin{figure}[ht]
\centering
\subfigure[
\label{fig:priornew}
Prior density in the parameters $(\log_{10}\umax,\log_{10}\smax)$. The thick red line represents the limit $\log\beta=0$, and the dotted lines mark $\log\beta=2,4,6,\ldots$
]{\includegraphics[width=0.5\textwidth]{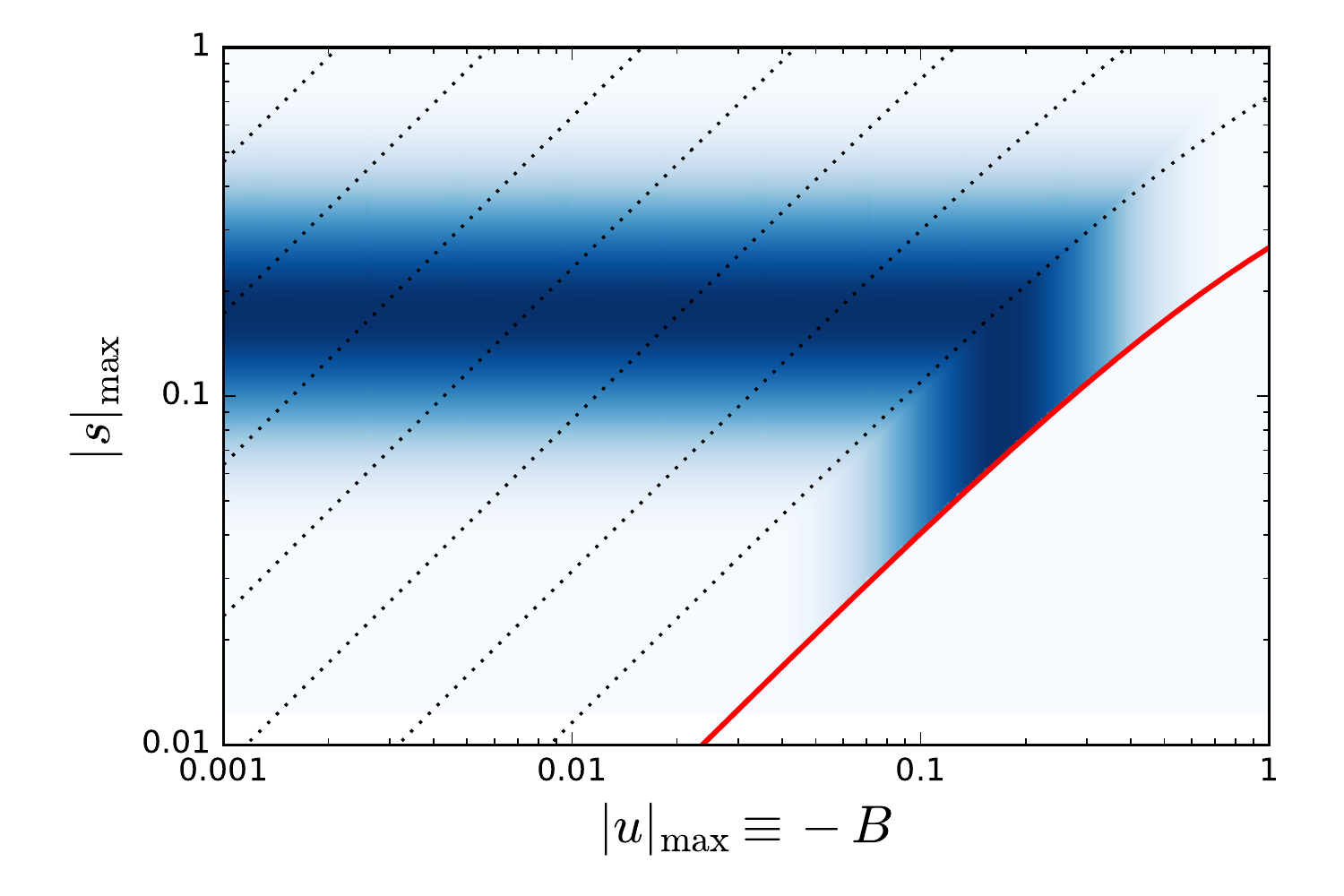}}
\subfigure[
\label{fig:priorold}
Region in the parameters $(B, \log\beta)$ sampled with \emph{uniform} prior density in our previous work \cite{Achucarro:2013cva,Achucarro:2014msa,Hu:2014hra}. The new prior density in figure \ref{fig:priornew} is plotted on top (shading), to highlight how much of the old prior is theoretically disfavoured.
]{\includegraphics[width=0.49\textwidth]{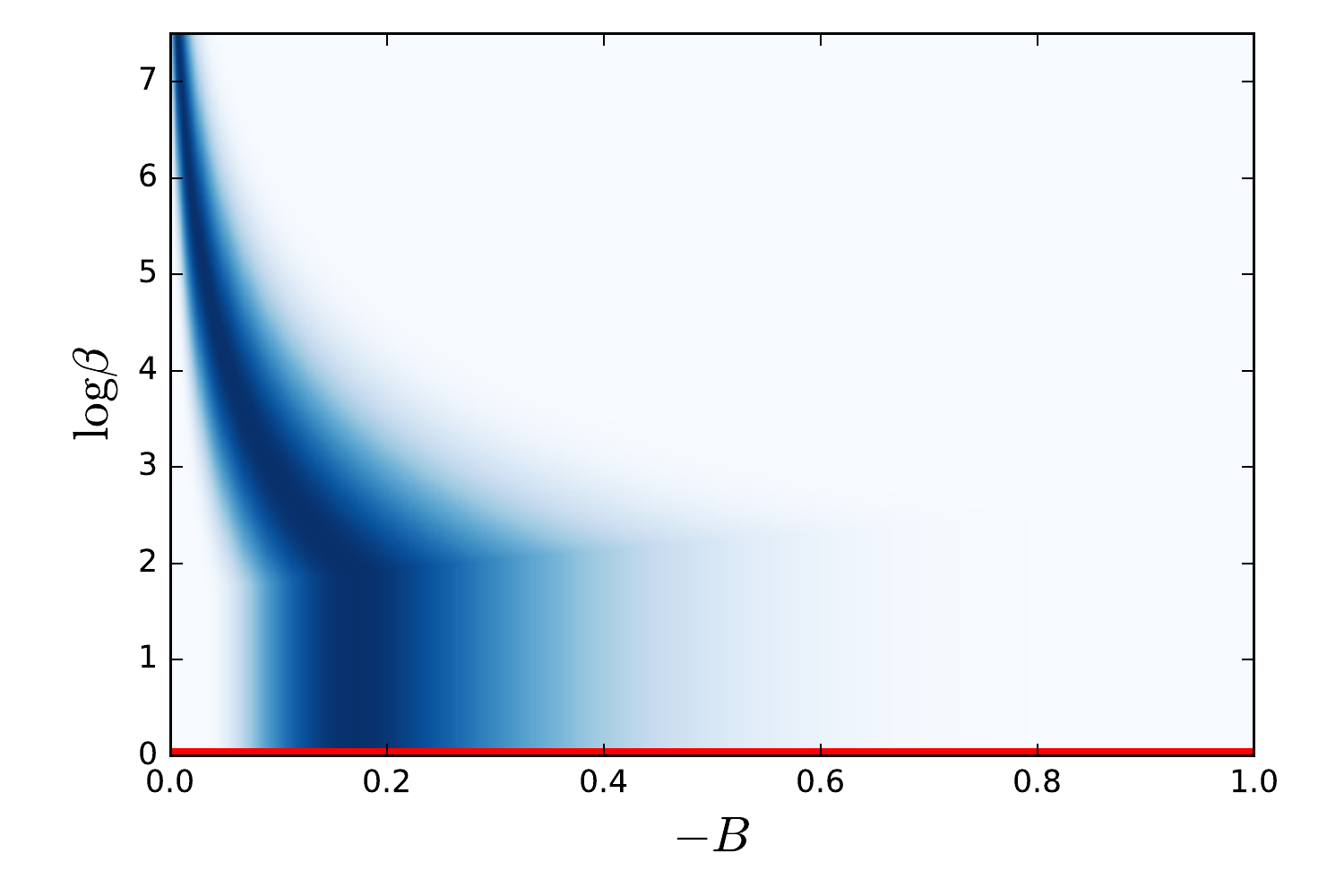}}
\caption{
\label{fig:prior}
Prior on intensity and sharpness of the speed of sound reduction in this work, \ref{fig:priornew}, and in previous works, \ref{fig:priorold}.}
\end{figure}

In this work, we make \textit{tabula rasa} and try to approach the prior choice in a model independent way, from the bare consistency requirements of the theoretical framework, eq.\ \eqref{eq:uslimits} in section \ref{sec:theoryreview}:
\begin{equation}
  \nonumber
  \max\left(\epsilon,|\eta|\right) \ll \max\left(\umax,\smax\right) \ll 1
\,.
\end{equation}
This condition \textit{on-the-maximum} gives the prior a \emph{framing square} shape, see figure \ref{fig:priornew}: above the diagonal \mbox{$\umax=\smax$}, the limits given by this equation must be imposed on $\smax$, whereas below the diagonal they must be imposed on $\umax$.


A good, simple choice for a prior that fulfils the condition above, translating the \emph{strong inequalities} into a probability density which softly falls towards the limits, would be a symmetric $\log$-Beta distribution\footnote{Not to be confused with the sharpness parameter $\beta$ of the Gaussian reduction defined above.} defined over the interval $[\max(\epsilon,|\eta|),\,1]$:\footnote{For a random variable $x$ in the domain $[0,1]$, $x\,\sim\,\Beta(a,b)$ has a probability density function $\mathcal{P}(x)=x^{a-1}(1-x)^{b-1}/N(a,b)$, with $N(a,b)\defeq \Gamma(a)\Gamma(b)/\Gamma(a+b)$.}
\begin{equation}
\max\left(\log_{10}\umax,\,\log_{10}\smax\right)
\,\sim\, \Beta(a,a)
\quad\text{with}\quad
a>1
,\quad
\max\left(\umax,\smax\right)\in[\max\left(\epsilon,|\eta|\right), 1]
\,.
\end{equation}
The lower the value of the shape parameter, $a>1$, the more disperse the distribution. The choice of a logarithmic pdf is based on the limits of the interval being typically two orders of magnitude apart. The symmetry of the $\Beta(a,a)$ distribution weighs both extremes equally, e.g.\ there is the same probability mass to the left of twice the lower limit, than to the right of half the upper limit. In this work, we choose $a=5$, which places the 95\% confidence level interval at approximately double/half the boundaries, and the 68\% at thrice/third.

The lower limit in the expression above depends on the slow-roll parameters $\epsilon$ and $\eta$. Strictly, we should impose a joint prior on $(\umax,\smax,\epsilon,\eta)$ which would account for the moving lower bound on eq.\ \eqref{eq:uslimits}. Alternatively, we could impose an equivalent prior on $(\umax,\smax,n_s,r)$, since $(n_s,r)$ are directly determined by the slow-roll parameters.


On the instant of maximum reduction $\tau_0$, the theoretical model imposes no requirements within the range $(\tau_i,0)$, where $\tau_i<0$ is the unknown conformal time at which slow-roll inflation started, and $\tau=0$ the conformal time corresponding to the end of slow-roll inflation. Regarding the density of the prior on $\tau_0$, two natural choices would be either a uniform prior on $\tau_0$ (no preference on the instant of maximum reduction \emph{in conformal time}) or a uniform prior on $\log(-\tau_0)$ (no preference \emph{in physical time} or in $\mathrm{e}$-folds, since $t\propto N\propto\log(-\tau)$). The choice depends on which of $t$ or $\tau$ one considers the natural time scale of inflation.


The prior distribution described above, for either choice of prior density on $\tau_0$, defines a weakly-informative Bayesian prior on the reduction of the speed of sound. It is independent of the particular model for the reduction, and motivated only by computational consistency.

Not all parameter combinations allowed by that prior generate features whose effect is observable within the CMB window of scales, and even among those that do, some are not easily distinguishable from a similarly looking change in the slow-roll or background parameters. We restrict ourselves to exploring the sub-space of the prior corresponding to features that are \emph{observable} and \emph{distinguishable}:
\begin{description}
\item[Observability:] If the reduction happens too early, it will not leave any trace on the observable scales of the CMB power spectrum window. One can easily check that, for reasonable values of $\umax$ and $\smax$, features happening before $\tau_0=-8000$ leave no trace in the CMB power spectrum. On the other hand, since $\umax=-B$ determines the amplitude of the feature in the power spectrum, we can ignore values of $\umax<10^{-3}$, which can never lead to significant improvements in the likelihood.\footnote{If we were fitting these features to the CMB bispectrum, we should allow for even smaller values of $\umax$, since the amplitude of the bispectrum features is also proportional to their sharpness, due to the derivatives in eq.\ \eqref{eq:bispSRFT}.}
\item[Distinguishability:] We discard parameter combinations corresponding to features whose appearance mimicks changes of the slow-roll parameters or the background cosmology. In \cite{Achucarro:2013cva}, we found that this was achieved by imposing that the feature is well contained within the observable scales, and that it performs at least four full oscillations within that window. Those conditions are guaranteed respectively by imposing a minimum sharpness of the Gaussian reduction of $\log\beta\ge0$ (see thick red line in figure \ref{fig:priornew}), and a minimum oscillatory frequency of $|\tau_0|\ge70$. This assumption also justifies ignoring effects from higher order slow-roll parameters, such as running of the spectral index.
\end{description}

This immediately defines the interesting range of $\tau_0$ to be explored: $\tau_0\in[-8000,-70]$. This interval covers three orders of magnitude, so the balance may be tilted towards a log-uniform choice. Nonetheless, we sample both choices, uniform and $\log$-uniform, to keep our analysis robust. In our previous work \cite{Achucarro:2013cva,Achucarro:2014msa,Hu:2014hra} and also here (see section \ref{sec:powerspectrum}), we find that $\tau_0$ is well constrained by the data, so the choice between priors here is not a vital one.


These assumptions also allow us to simplify the prior on $(\umax,\smax)$. The requirements for \emph{distinguishability} ensure that there are no significant degeneracies in the posterior between the feature and the slow-roll parameters, i.e.\ the estimation of the slow-roll parameters from Planck data are robust with respect to the introduction of the feature. This robustness means that we can fix the lower limit in eq.\ \eqref{eq:uslimits} to the values found by Planck for the slow-roll parameters: although relaxing that limit would allow for smaller values of the feature parameters, those would never produce significant posterior probability, since they necessarily correspond to disfavoured values of the slow-roll parameters.

We choose to fix that lower bound in eq.\ \eqref{eq:uslimits} to the \emph{central value} of Planck's estimate for $\eta$ (the largest of the slow-roll parameters), using temperature and polarisation data and assuming a featureless power spectrum with free running of the spectral index \cite{Ade:2015lrj}. That is $\eta=0.03$. The choice of the central value instead of the upper bound is not necessarily problematic, since the prior density on $\max(\umax,\smax)$ decays fast towards that limit: for a $\Beta(5,5)$, Planck's 2-$\sigma$ upper bound $\eta\approx0.05$ falls under the leftmost $5\%$ of prior mass of $\max(\umax,\smax)$.


In summary, the sub-space of the Bayesian prior that we actually explore is given by
\begin{equation}
\tau_0\in[-8000,\,-70]
\qquad\text{and}\qquad
\max\left(\log_{10}\umax,\,\log_{10}\smax\right)
\,\sim\, \Beta(5,5)
,\quad
\max\left(\umax,\smax\right)\in[0.03,\,1]
\,,
\end{equation}
with either a uniform or a log-uniform density on $\tau_0$ and additional limits on $(\umax,\smax)$ given by $\log\beta\ge 0$ and some minimum value for $\umax$ for which the features would be unobservable in the power spectrum due to their small intensity ($\umax\ge10^{-3}$ would be enough; in practice, we use $\log\beta\le14$ for this limit, imposing a minimum value for $\umax$ in the range $10^{-4}$--$10^{-3}$, depending on $\smax$). The density of this prior corresponds to the shading in figure\ \ref{fig:priornew}.

We shall not forget that the regions of the full prior discarded by observability and distinguishability are actually allowed by the theory, and therefore the full prior must be taken into account in a full evidence computation. But such computation is beyond the scope of this paper.

Let us now compare the new prior with the one we used in our previous work \cite{Achucarro:2013cva,Achucarro:2014msa,Hu:2014hra}, which is uniform over the region plotted in figure \ref{fig:priorold}. If we plot the density of the new prior on top of the old uniform prior, we see that approximately $\sfrac{2}{3}$ of its area is unshaded, i.e.\ has null probability density under the new prior. If we trust that the new prior appropriately accounts for the consistency requirements of the theory, then sampling from the old prior leads to over-sampling theoretically uninteresting regions, while under-sampling the interesting ones. Thus, we consider the present choice more reasonable and efficient, since not only are we more likely to find fits of theoretically allowed features, but we are also able to do it in a fraction of the sampling time.

\section{Data sets and sampling methodology for the power spectrum}
\label{sec:methodology}

The features from the reduction are computed using a \textit{fast Fourier transform} to perform the integral of the reduction in eq.\ \eqref{eq:dpp}. The primordial power spectrum is then fed to a modified version of the \texttt{CAMB} Boltzmann code \cite{Lewis:1999bs,Howlett:2012mh} (\mbox{\url{http://camb.info}}). We modified \texttt{CAMB} to adaptively increase the sampling density on $k$ and $\ell$ only where necessary.

The features are fitted to the \emph{unbinned} CMB TT, TE and EE power spectra of the Planck 2015 data release \cite{Adam:2015rua,Aghanim:2015xee}. The inclusion of the polarised spectra is {an update on} the previous searches that we performed using Planck's 2013 data \cite{Achucarro:2013cva,Achucarro:2014msa}. The use of the unbinned likelihoods is justified by the high oscillatory frequency that the features can reach: the $\Delta\ell=30$ binning of the multipoles corresponds roughly to a binning of $\Delta k=2\times10^{-3}\,\mathrm{Mpc^{-1}}$ in primordial scales, which is smaller than a full oscillation as soon as $|\tau_0| > 1500$, and we do explore much higher values.

The sampling is performed with the sampler/integrator \texttt{PolyChord} \cite{2015MNRAS.453.4384H}, which was chosen especially because of its multi-modal sampling capabilities, since we know the likelihood to be multimodal from previous searches \cite{Achucarro:2013cva,Achucarro:2014msa}. Handling of the theory and likelihood codes and the sampler is performed with \texttt{CosmoChord}, a modified version of \texttt{CosmoMC} \cite{Lewis:2002ah} that incorporates \texttt{PolyChord} as a sampler.\footnote{Our last search \cite{Hu:2014hra} was conducted with the \texttt{MultiNest} nested sampling algorithm \cite{Feroz:2007kg,Feroz:2008xx,Feroz:2013hea}. The \texttt{PolyChord} sampler used in this work is an improvement on \texttt{MultiNest}, that it is tailored for high-dimensional parameter spaces, thanks to the use of slice sampling at each iteration to sample within the hard likelihood constraint of nested sampling.}

For the sake of performance, the value of the nuisance parameters of the Planck 2015 likelihood, which describe the foreground effects and experimental calibration that affect the CMB measurement, are fixed to their best fit achieved by the Planck Monte Carlo sample with binned, polarised baseline likelihood (\texttt{lowTEB + plikHM\_TTTEEE}) and baseline $\Lambda$CDM model.\footnote{{See table 2.6 in \protect\url{https://wiki.cosmos.esa.int/planckpla2015/images/f/f7/Baseline_params_table_2015_limit68.pdf}.}}
\footnote{We have verified that varying the nuisance parameters has a negligible effect on our results.}
When not sampled (e.g.\ in the bispectrum study), the cosmological parameters are fixed to the best fit of that same sample.

For each choice of prior density for $\tau_0$, we have run \texttt{CosmoChord} with 16 MPI processes, each allowed to thread across 8 CPU cores. The \texttt{PolyChord} algorithm has been run in multi-modal mode, with $1000$ live points, and a stopping criterion of $1/100$ of the total evidence contained in the final set of live points. Since we have fixed the value of the nuisance parameters, there was no speed hierarchy of which to take advantage. With these parameters, each run was completed within a few days.

\section{Results of fits to the power spectrum}
\label{sec:powerspectrum}

We have performed the sampling on the CMB power spectrum data as described in the last section, varying the baseline $\Lambda$CDM cosmological parameters $(\Omega_\mathrm{b}h^2,\,\Omega_\mathrm{c}h^2,\,\theta_\mathrm{MC},\,\tau_\mathrm{reio},\,\log A_\mathrm{s},\,n_\mathrm{s})$ over a wide uniform prior, and the feature parameters $(\tau_0,\,\umax,\,\smax)$ using the prior described in section \ref{sec:prior}.

\begin{figure}[p]
\centering
\subfigure[
\label{fig:trilog}
]{\includegraphics[width=0.6\textwidth]{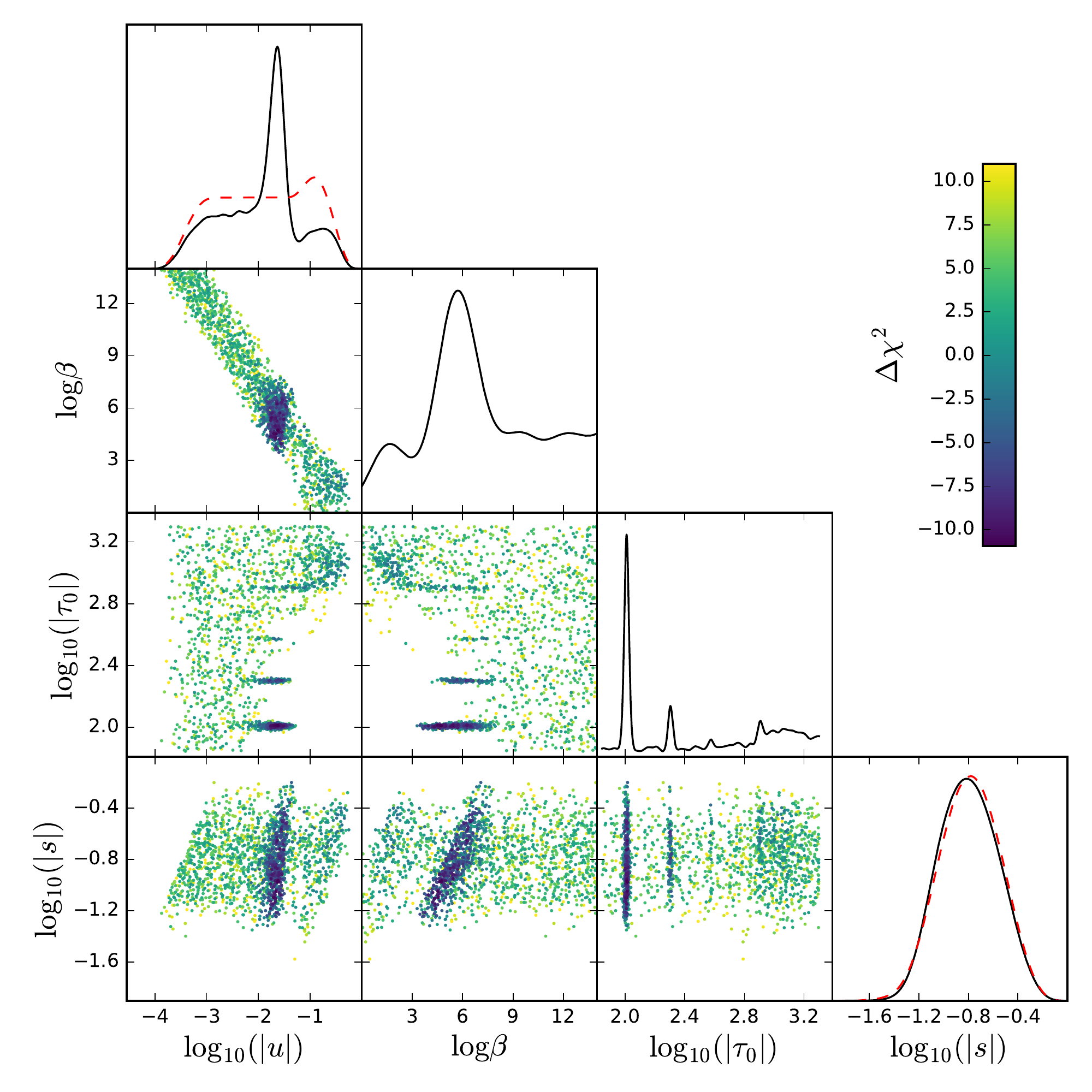}}
\subfigure[
\label{fig:trilin}
]{\includegraphics[width=0.6\textwidth]{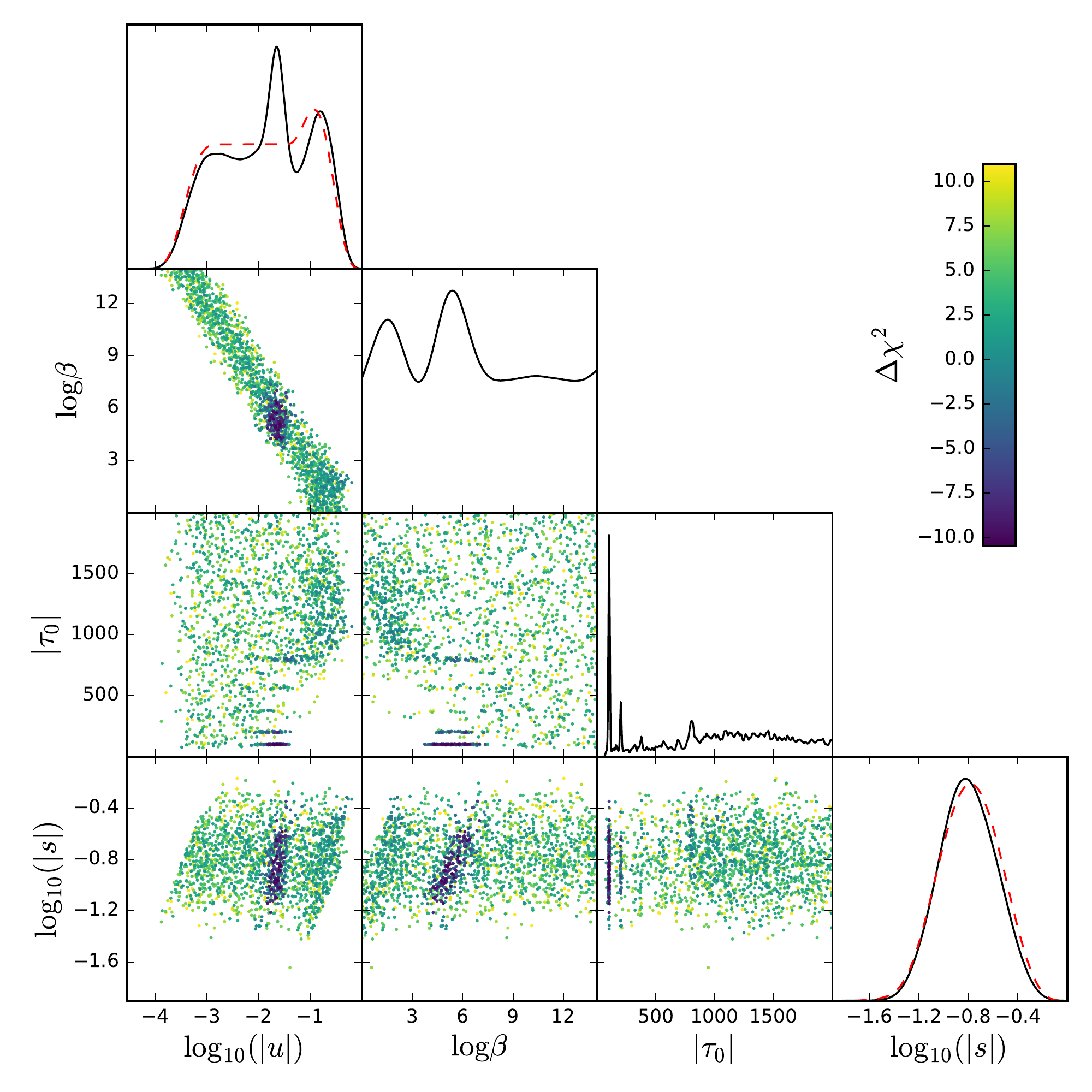}}
\caption{
\label{fig:tri}
{
Marginalised 1d posteriors and 2d $\chi^2$ scatter plots for the feature parameters $(\tau_0,\umax,\smax)$ and the derived parameter $\log\beta$ of eq.\ \eqref{eq:gaussN}. The prior on $(\umax,\smax)$ is described in section \ref{sec:prior}, with the 1d marginalised prior in dashed red line -- compare the $(\umax,\smax)$ posterior with the prior in figure \ref{fig:prior}, and notice how almost nothing is learnt about $\smax$ from the data. The prior on $\tau_0$ is either log-uniform, \ref{fig:trilog}, or uniform, \ref{fig:trilin}. The colour scale shows the $\Delta\chi^2$ of the unbinned, polarised Planck 2015 likelihood, and differences are given with respect to the best fit of the baseline model to a featureless power spectrum. The modes observed along $\tau_0$ are described in table \ref{tab:maps}.}
}
\end{figure}

As stated in section \ref{sec:prior}, we have sampled \emph{twice}, with two different priors for $\tau_0$: one is a uniform prior on $|\tau_0|$, which assigns equal probability for a reduction occurring at any \emph{conformal} time, and another with a uniform prior in $\log_{10}|\tau_0|$, which assigns equal probability for a reduction occurring at any \emph{physical} time. Both cases are physically well motivated. The result of both samples can be seen in figure \ref{fig:tri}, and the most relevant modes are shown in table \ref{tab:maps}. The reference value $\chi^2=34655.5$ for the effective $\chi^2$, used in figure \ref{fig:tri} and table \ref{tab:maps}, has been obtained from a run with the same likelihood and a featureless primordial power spectrum. Those differences in $\chi^2$ are shown as an approximate reference, since we have not used a thorough maximisation algorithm. The size of the decrease in $\chi^2$ does not amount to a detection, neither does the Bayes ratio: this model is disfavoured with respect to the baseline $\Lambda$CDM when considering power spectrum data only.

We found no significant degeneracies between the parameters of the feature and those of the baseline cosmological model; the correlation coefficients stay below $|\rho|<0.1$ for most combinations, and only for some combinations with $(\Omega_\mathrm{b}h^2,\,\Omega_\mathrm{c}h^2,\,n_\mathrm{s})$ the correlation coefficient grows up to $|\rho|\le0.18$, which is still smaller than what was found in fits to the 2013 data \cite{Achucarro:2014msa}. This is consistent with the assumptions made in section \ref{sec:prior} in order to avoid those degeneracies, namely the lower bounds $|\tau_0|\ge70$ and $\log\beta\ge0$, which together enforce a minimum number of oscillations to occur within the CMB window.

\begin{table}[t]
\centering
\begin{tabular}{c|ccc|cc|c}
Mode name & $|\tau_0|$ & $\log_{10}\umax$ & $\log_{10}\smax$ & $-10^{2}\,B$ & $\log\beta$ & $\Delta\chi^2_\mathrm{MAP}$ \\
\hline\hline
\textit{100} & \three{100}{102}{105}& \three{-1.82}{-1.59}{-1.47}&
   \three{-1.11}{-0.91}{-0.62} & \three{1.5}{2.6}{3.4} & \three{4.4}{4.8}{6.8} & -11\\
\textit{200} & \three{195}{203}{207}& \three{-2.16}{-1.62}{-1.44}&
   \three{-1.01}{-0.78}{-0.57} & \three{0.7}{2.4}{3.7} & \three{4.5}{5.6}{8.2} & -8\\
\hline
\textit{800} & \three{770}{801}{830}& \three{-2.06}{-1.37}{-0.77}&
   \three{-0.94}{-0.53}{-0.47} & \three{0.1}{4.3}{17.0} & \three{2.0}{5.6}{8.2} & -6\\
\textit{1000} & \three{935}{1099}{1631} & \three{-2.78}{-0.51}{-0.45} &
   \three{-1.03}{-0.63}{-0.54} & \three{0.1}{31}{35} & \three{0}{1.5}{7.2} & -4.5
\end{tabular}
\caption{\label{tab:maps}
{
68\% confidence level intervals and \textit{maxima a posteriori} (MAP, in parenthesis) for the modes described in the text and visible in figure \ref{fig:tri}. The $\Delta\chi^2$ of the MAP's are given with respect to the best fit of the baseline model to a featureless power spectrum. We also provide c.l.\ intervals for the derived Gaussian ansatz parameters $(B,\log\beta)$. The c.l.\ intervals of $\tau_0$ for modes \textit{100} and \textit{200} correspond to a Gaussian in $\log|\tau_0|$. Notice the similarity of the bounds on $\smax$ along the table: they all correspond approximately to the prior limits (sec.\ \ref{sec:prior}).}
}
\end{table}

Looking at the marginalised posterior for $|\tau_0|$, we identify {the following modes (see table \ref{tab:maps}):}
\begin{description}
\item[Low $\boldsymbol{|}\boldsymbol\tau_{\boldsymbol0}\boldsymbol{|}$:]
{
two modes at $\tau_0\sim-100,-200$. They both have a very well-determined oscillation frequency $\tau_0$ and an amplitude $\umax$ of a few $0.01$'s. Due to their low $\tau_0$, we take their confidence level intervals from the log-uniform-$\tau_0$ sample, where they are better resolved. Both modes correspond to the sharp regime, $\smax\gg\umax$, or high $\beta$.}
\item[High $\boldsymbol{|}\boldsymbol\tau_{\boldsymbol0}\boldsymbol{|}$:]
{
one mode at $\tau_0\sim-800$, with characteristics similar to the two modes above, but worse $\chi^2$ and looser constraints on the parameters. Also, a much broader mode with $\tau_0\sim-1100$, with a wide posterior on $\tau_0$, unbounded amplitude (constrained by the prior), and a clearly lower sharpness $\beta$ than the rest of the modes, which places it in the not-so-sharp regime, $\umax\approx\smax$. Despite their different regime, the boundary between these two modes is not clearly defined, so we have imposed it at $\tau_0=-840$ -- thus their 68\% c.l.\ limits on $\tau_0$ are just an approximation.
}
\end{description}

On the very high $|\tau_0|>2000$ region, we do not find any significant mode. This is probably due to their high oscillatory frequency: the transfer functions are almost constant with respect to them, so their projection on the CMB sky smears out most of their intensity, needing too high values of $\umax=|B|$ that are disfavoured by the prior.

In all the modes above, $\smax$ is constrained by the prior only. This lack of predictivity on $\smax$ was already observed in our previous work with Planck 2013 data \cite{Achucarro:2013cva,Achucarro:2014msa,Hu:2014hra}; {there, it appeared as} a degeneracy between the parameters $(B,\,\log\beta)$ of the Gaussian ansatz of eq.\ \eqref{eq:gaussN}. Moving along that degeneracy eventually saturated the $\smax<1$ bound, which is avoided now by the new and more realistic prior. That degeneracy still persists, in a milder version, between $\log_{10}\umax$ and $\log_{10}\smax$. {As explained in 
\cite{Achucarro:2013cva,Achucarro:2014msa}, the degeneracy was caused by the fact that a simultaneous increase in $|B|$ and $\log\beta$ produces almost no changes in the aspect of the feature in the CMB power spectrum \cite[figure 9]{Achucarro:2014msa}: a larger $\log\beta$ shifts the mode towards smaller scales, where damping and lensing erases most of the primordial information, while a larger value of $|B|$ keeps the power at larger scales constant. The new prior avoids this effect, as it is illustrated by the difference between the current $(\tau_0,\log\beta)$ profile in figure \ref{fig:tri} and the corresponding ones in our previous work: \cite[figure 1]{Achucarro:2013cva}, \cite[figure 5]{Achucarro:2014msa} and \cite[figure 2]{Hu:2014hra} -- in the last ones, the mode continues towards higher values of $\log\beta$ with almost constant $\chi^2$, well past the $\smax=1$ mark.}

Comparing these results with our previous searches in Planck 2013 and WiggleZ data \cite{Achucarro:2013cva,Achucarro:2014msa,Hu:2014hra}, we see that the modes at $\tau_0\approx-100,-200$ correspond respectively to the modes $\mathcal{E},\,\mathcal{C}$ already found there.\footnote{Notice that mode $\mathcal{E}$ on Planck 2013 was previously discarded due to its low $\umax$ and $\smax$. The corresponding one in Planck 2015 does not have that problem, since at least $\smax$ is large enough.} Mode $\mathcal{A}$ appears as a very faint mode {with $\tau_0\approx-377$ and $\umax\approx0.02$}. However, modes $\mathcal{B}$ and $\mathcal{D}$ of Planck 2013 have no corresponding significant signal in Planck 2015, neither does the mode found at $\tau_0\approx-540$ in \cite{Hu:2014hra}. {To check whether those modes are still present in Planck 2015 but have been suppressed by the new prior,} we re-ran the chains with the old non-realistic prior, uniform on $(B,\log\beta)$, and the binned likelihoods -- we still found no trace of 2013's modes $\mathcal{B}$ or $\mathcal{D}$, but we did find a mode close to $\tau_0\approx-540$, albeit with a very high $\smax$ that would discard it under the new prior. The disappearance of mode $\mathcal{B}$ may be related to that mode's benefiting from the spurious wiggle at $\ell\sim1800$ in Planck 2013's TT power spectrum.

To assess the effect of the new high-$\ell$ CMB polarisation data in our samples, we repeated the analysis of the uniform-$\tau_0$ case with Planck 2015's unbinned TT power spectrum likelihood plus the low-$\ell$ polarised likelihood. We found that the high-$\ell$ polarised data enhances the mode \textit{100} while it significantly dampens the mode \textit{1000}, which shows up more intensely and with a sharper $\tau_0$ c.l.\ interval when using TT+lowTEB. The other two modes do not receive a large correction.

\begin{figure}[ht]
\centering
\includegraphics[width=0.80\textwidth]{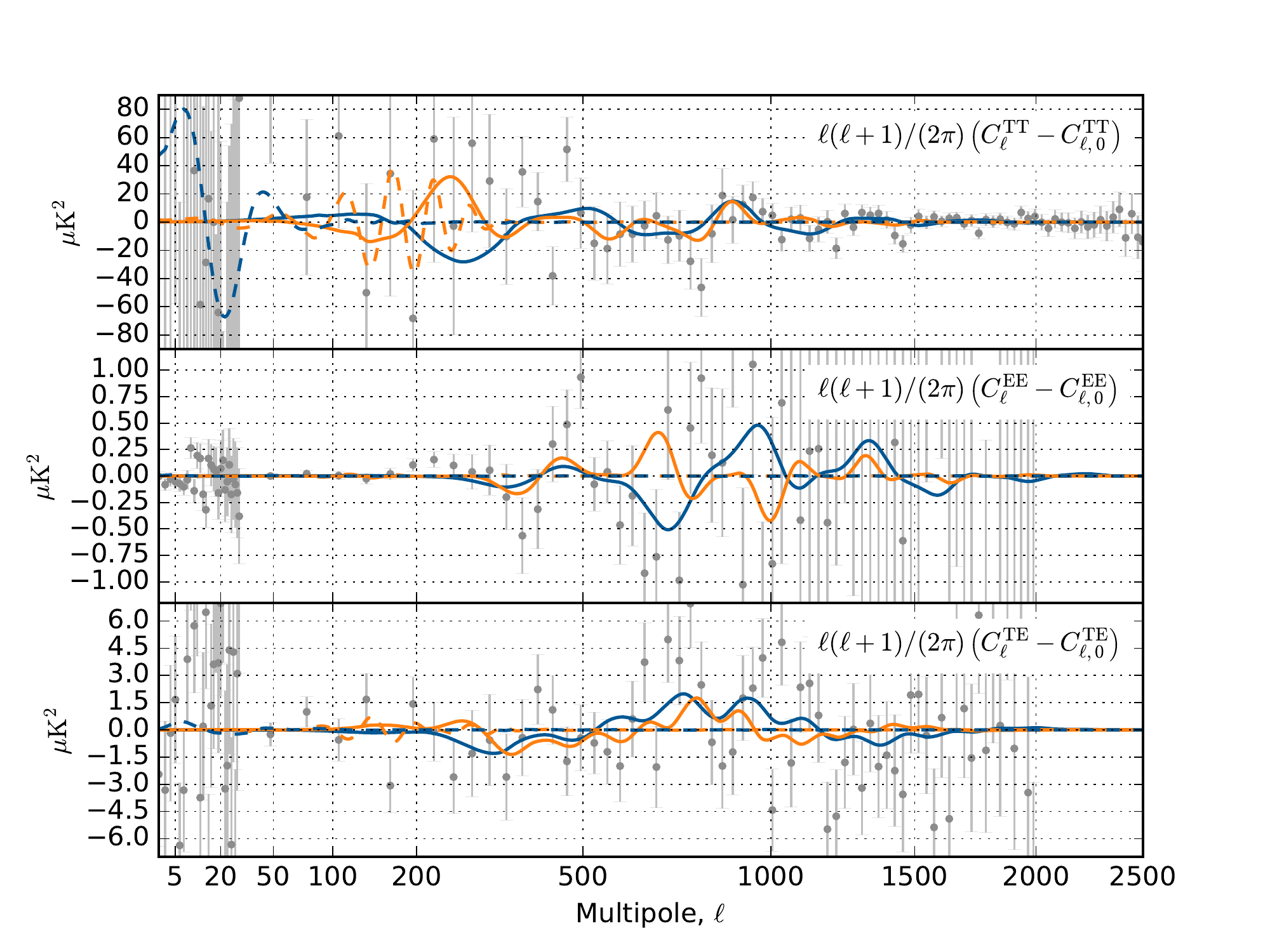}
\caption{
\label{fig:Cl}
Differences between the best fit to the Planck 2015 power spectrum (using polarised low- and high-$\ell$ likelihoods \cite{Ade:2015xua}) of the $\Lambda$CDM baseline model, and the MAP's of modes \textit{100} (blue solid, darker), \textit{200} (orange solid), \textit{800} (orange dashed) and \textit{1000} (blue dashed, darker) from table \ref{tab:maps}. Notice how mode \textit{1000} (blue dashed, darker) fits the minus/plus structure at $\ell\approx20$--$40$, how mode \textit{800} (orange dashed) fits some apparent wiggles at the fist acoustic peak, and how modes \textit{100} and \textit{200} fit small deviations from the baseline model across a higher range of multipoles (cf.\ figure \ref{fig:per}).}
\end{figure}

\begin{figure}[t]
\centering
\includegraphics[width=0.8\textwidth]{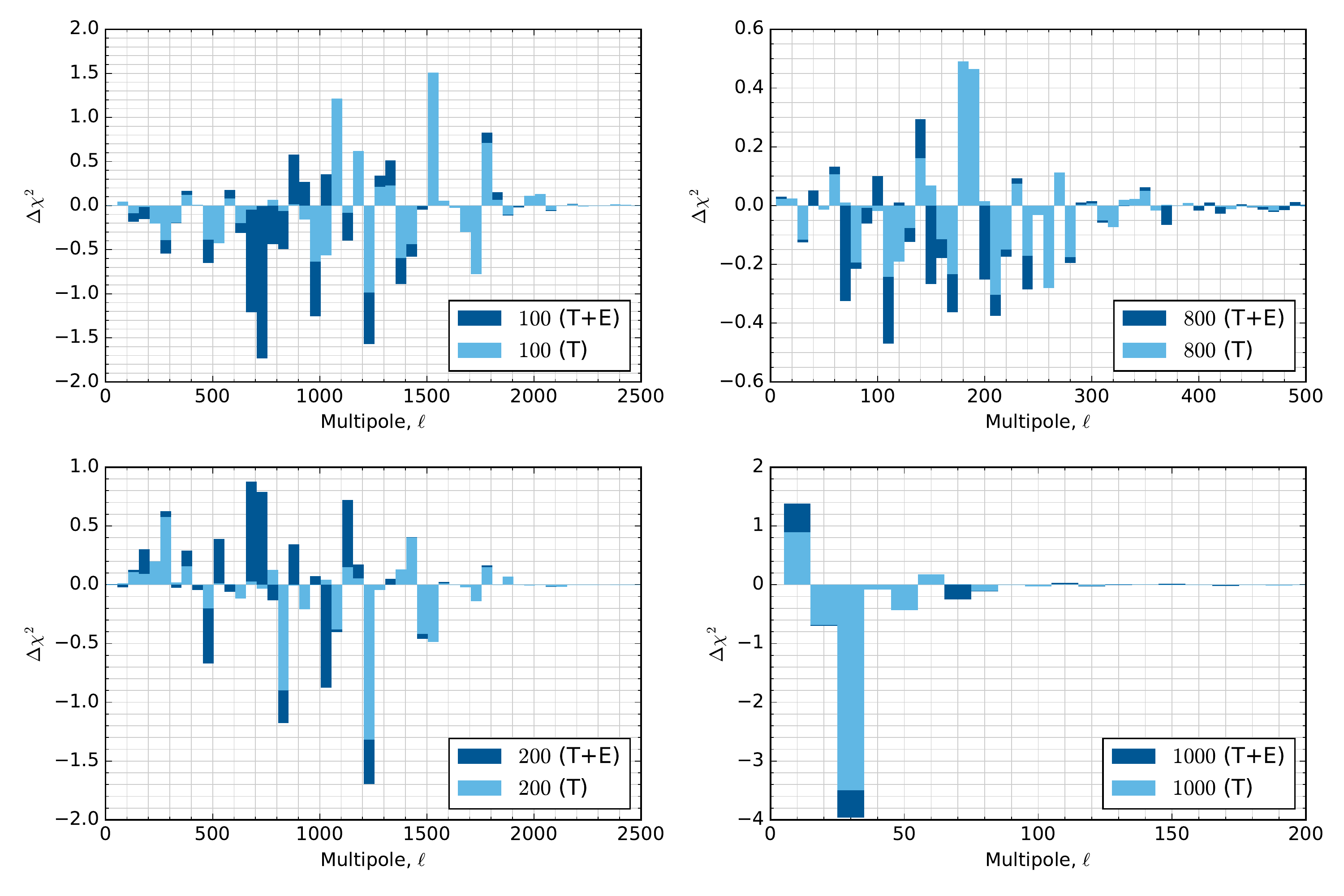}
\caption{
\label{fig:per}
Difference in effective $\chi^2$ per multipole between the maxima a posteriori of the modes cited in table \ref{tab:maps} and the best fit to the Planck 2015 power spectrum of the $\Lambda$CDM baseline model, using temperature and polarisation data (dark) and temperature data only (clear), cf.\ figure \ref{fig:Cl}. Negative values indicate a better fit by the feature model. For the sake of clarity, the multipoles are approximately binned proportionally to the oscillation frequency of each mode.}
\end{figure}

The residuals of these modes with respect to the best fit of a featureless $\Lambda$CDM baseline model are shown in figure \ref{fig:Cl}, and their respective improvements in goodness of fit $\chi^2$ per multipole are shown in figure \ref{fig:per}. We can see that modes \textit{100} and \textit{200} span across most of the multipole range, fitting diverse structures in TT and EE. The mode \textit{800} is restricted to the first acoustic peak and fits a small number of apparent wiggles in the data. The \textit{maximum a posteriori} (MAP) of \textit{1000} tries to fit the dip in temperature at $\ell\approx 20$ and the following peak at $\ell\approx 40$, at the cost of raising the power at $\ell\approx 10$ and below (a similar feature in 2015 data has been reported in \cite{Ade:2015lrj,Cai:2015xla,Benetti:2016tvm,Hazra:2016fkm,GallegoCadavid:2016wcz}). Despite the goodness of the fit, this is in conflict with the apparent lack of power at very small multipoles seen in Planck's data, and may impose an even more stringent upper limit on the relative amplitude $r$ of the tensor primordial power spectrum. We leave the study of this possibility for future work.

\begin{figure}[ht]
\centering
\includegraphics[width=0.5\textwidth]{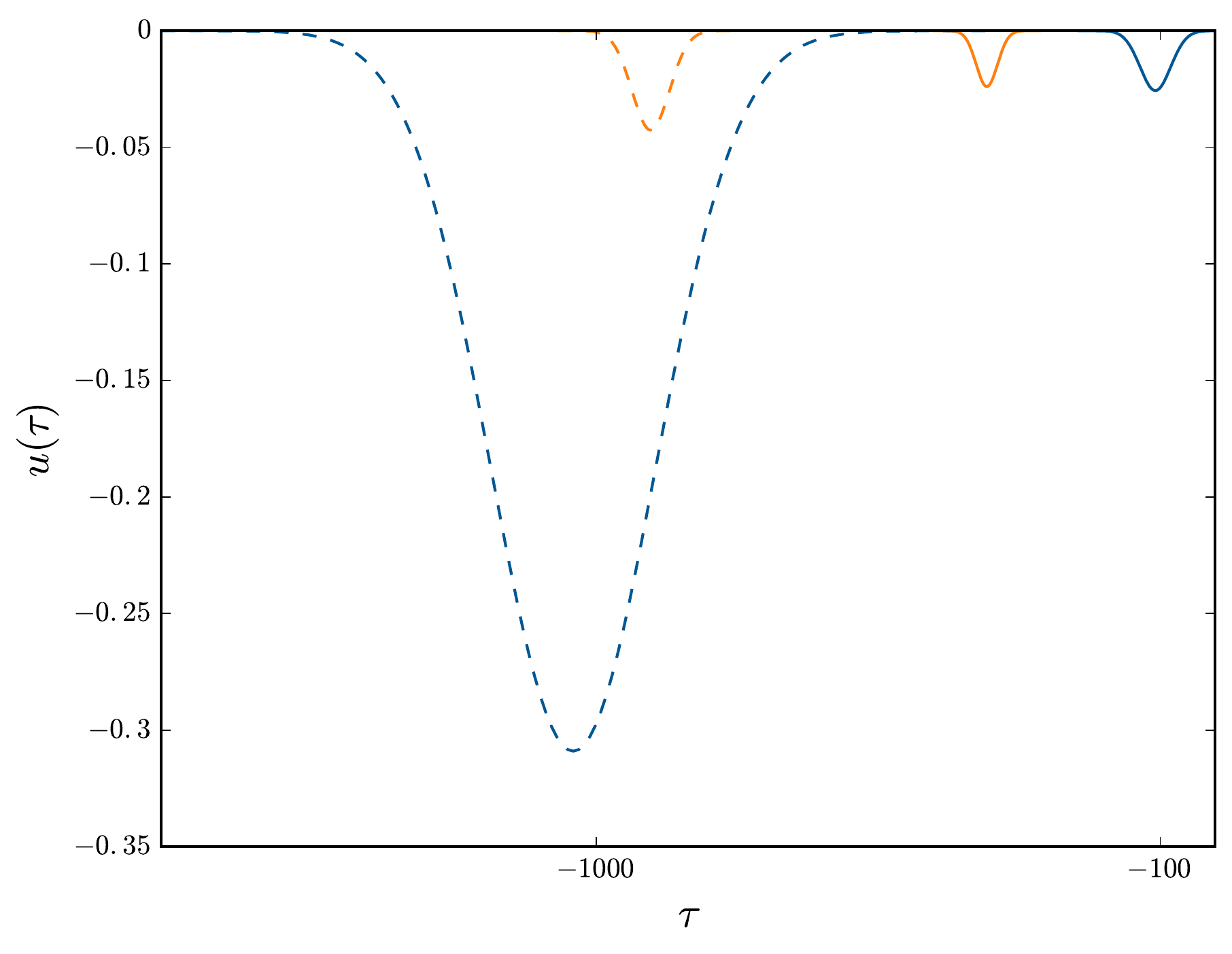}
\caption{
\label{fig:u}
Speed of sound reduction in terms of $u=1-\cs^{-2}$ for the modes described in table \ref{tab:maps}, in logarithmic scale for $\tau$, and with the same colours as in figure \ref{fig:Cl} (the correspondence between colours and $\tau_0$'s is here obvious).}
\end{figure}

We could ask whether two or more of those modes could be present in the data simultaneously. This would correspond to the case of the inflaton suffering two consecutive reductions in its sound speed, e.g.\ due to two consecutive turns in field space. The complete answer to that question would come from a fit of two simultaneous features with the restriction that they do not overlap in $\tau$. Their respective features in the CMB power spectrum may or may not overlap. The subset of the parameter space in which the power spectrum features do not overlap can be characterised using the present search: any pair of the modes that we found that do not overlap either on $\tau$ or on the power spectrum could have occurred together. Looking at figures \ref{fig:Cl} and \ref{fig:u}, we observe that the only two possible combinations would be those of mode \textit{1000} with either \textit{100} or \textit{200}. 

\section{Predicted bispectrum features}\label{sec:bispectrum}

\begin{figure}[ht]
\centering
\includegraphics[width=0.7\textwidth]{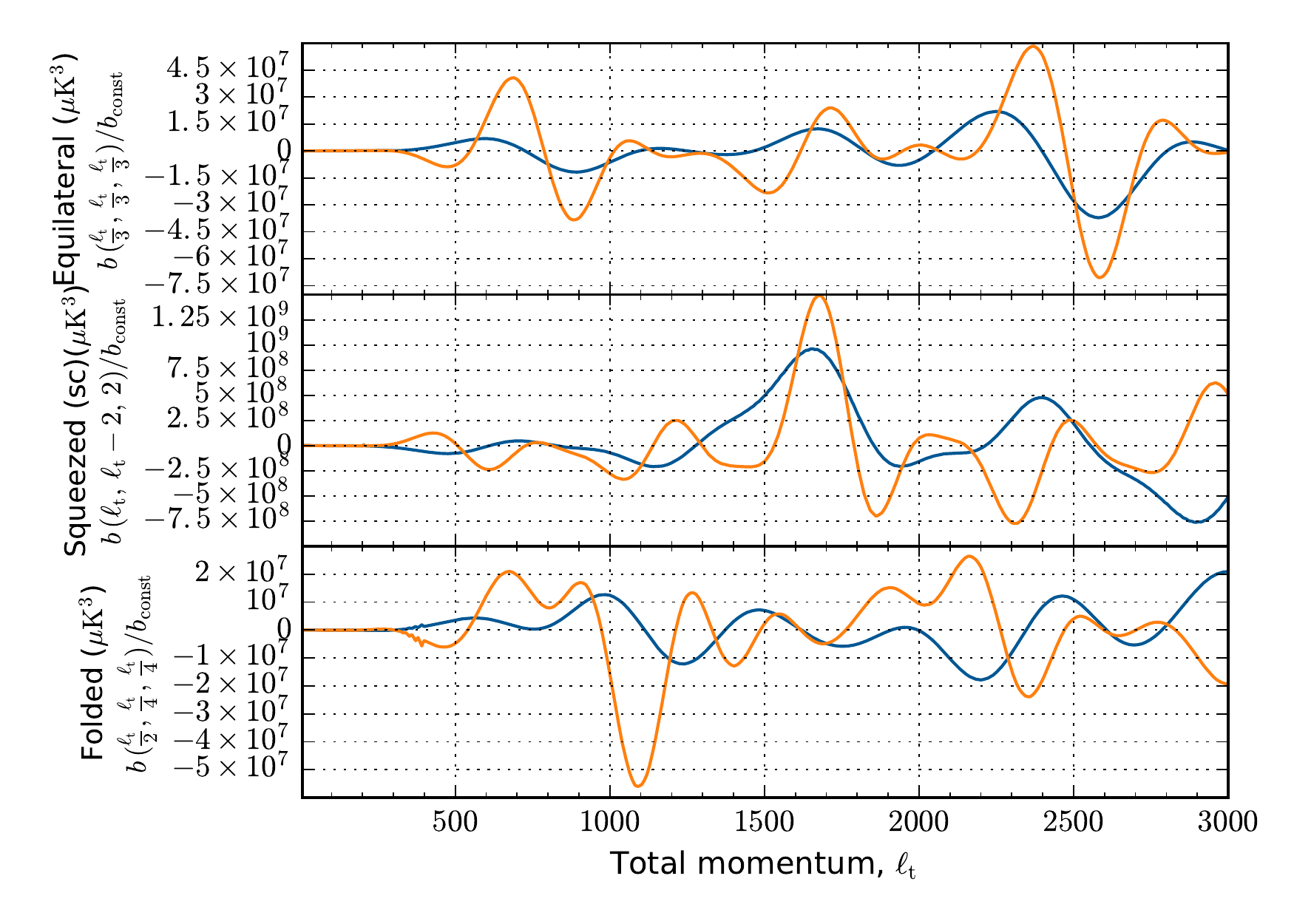}
\caption{
\label{fig:cmbB100200}
Equilateral, squeezed and folded limits of the CMB TTT bispectrum of the \textit{maximum a posteriori} of modes \textit{100} (blue, darker) and \textit{200} (orange). The x axis is the total scale $\ell_\mathrm{t}\defeq\ell_1+\ell_2+\ell_3$ and the bispectra are weighted by a constant shape in the Sachs-Wolfe approximation: $b_\mathrm{const} = (27\prod_{i=1}^3(2\ell_i+1))^{-1}((\ell_\mathrm{t}+3)^{-1}+\ell_\mathrm{t}^{-1})$  \cite{Fergusson:2008ra,Liguori:2010hx,Fergusson:2010dm}. In the equilateral limit, notice how both bispectra present a plus-minus structure in the first acoustic peak ($\ell\in[400, 1200]$) and a negative peak at $\ell_\mathrm{t}\approx 2600$, similar to what is described on Planck TTT bispectrum data~\cite{Ade:2015ava}.}
\end{figure}

We have computed the CMB temperature bispectrum (TTT) (see e.g.\ figure \ref{fig:cmbB100200}) using an extension of the expansion in total scale proposed in \cite{Munchmeyer:2014cca}, described in appendix \ref{app:bispectrum}. As expected, and similarly to what happens for the power spectrum, we find the CMB TTT bispectrum to be close to the primordial oscillatory shape described in sec.\ \ref{sec:theory} and figure \ref{fig:primordialB}, modulated by the transfer functions.

Due to the the lack of a publicly released bispectrum likelihood, we have not been able to perform a joint analysis of the power spectrum and bispectrum. But already at this point, we can use the posterior modes in the last section (see table \ref{tab:maps}) to make predictions for future searches in the bispectrum and to compare them to present searches of similar templates, if any. The basis of those predictions is the narrow constraints on the oscillatory frequencies $|\tau_0|$ of the features, and their rather well-defined intensity $\umax$, specially for modes \textit{100} and \textit{200}, but also for mode \textit{800} to a lesser extent. If a fit to the bispectrum of this kind of features hits any of these thin regions in $\tau_0$ and shows a similar intensity $\umax$, this would strongly hint towards the presence of a reduction in $\cs$ in the regime considered here. As we stated in the previous section, the power spectrum data is not able to constrain $\smax$ beyond its prior. Thus, we cannot predict a more concrete value for it.

We would like to especially remark modes \textit{100} and \textit{200} (see table \ref{tab:maps}) as predictions for a signal in the bispectrum. Their TTT bispectra (see figure \ref{fig:cmbB100200}) approximately presents some of the characteristics described in the reconstructed Planck bispectrum (sec.\ 6.2.1 in \cite[sec.\ 6.2.1]{Ade:2015ava}: a plus-minus structure in the equilateral limit at the $\ell$'s corresponding to the first acoustic peak, $\ell_\mathrm{total}\in[400,1200]$, and a negative peak (though preceded by a positive one) associated to the equilateral third acoustic peak, $\ell_\mathrm{total}\in[2300,3000]$. They also present additional structure in other limits and scales, where nothing has been particularly reported by the Planck collaboration in~\cite{Ade:2013ydc,Ade:2015ava}, except for a mention of small features in the folded limit and the squeezed limit, the last one claimed to be associated to the ISW-lensing secondary bispectrum. The coincidence between the bispectra predicted by modes \textit{100} and \textit{200} and the description of Planck 2015's bispectrum is tantalising, given that the predicted features come from a fit to the power spectrum only.

We can assess the likelihood that our predictions are found when tested directly on Planck 2015 data, as well as their correlations with other primordial and secondary templates that have already been searched for. To do that, we use the Fisher matrix formalism, assuming an idealised version of Planck's effective beam and noise, and taking into account the TTT bispectrum only. In this formalism, the signal-to-noise ratio of a bispectrum template under an experimental model is given by the square root of the auto-correlation of the template through a covariance matrix that accounts for the expected experimental errors; the correlation between two templates, carrying the meaning of the fraction of the intensity of a template that can be inferred from a measurement of a different one, is given by the covariance between these two templates. The details of how this signal-to-noise and correlations have been computed can be found in appendix \ref{sec:fisher}. The results for modes \textit{100} and \textit{200} are shown in table \ref{tab:corrs}. The signal to noise for both modes is approximately the same, and it could possibly grant a detection if this template was tested against data containing the corresponding physical signal. We can also observe that between one half and two thirds of the signal-to-noise stems from the divergent squeezed limit only. We have also computed the correlation between modes \textit{100} and \textit{200} and the ISW-lensing bispectrum, and found it to be very small despite the oscillatory nature in the squeezed limit of both the ISW-lensing bispectrum and our template.

\begin{table}
\centering
\begin{tabular}{c|c|c|c|c|c}
~~Mode~~ & ~~S/N~~ & ~S/N$_\mathrm{squeezed}$~ & ~~Corr $\cos$~~ & ~~Corr $\sin$~~ & ~~Corr ISW-l~~ \\
\hline\hline
\textit{100} & $7.4$ & $4.5$ & $-0.26$ & $-0.59$ & $-0.03$ \\\hline
\textit{200} & $7.5$ & $4.0$ & $-0.21$ & $-0.65$ & $-0.04$ \\\hline
\end{tabular}
\caption{
\label{tab:corrs}
Signal-to-noise in the TTT bispectrum of the \textit{maxima a posteriori} (MAP) of modes \textit{100} and \textit{200} and their isolated respective squeezed contributions. The signal-to-noise is referred to the amplitude $\umax$ of the MAP for each mode, i.e.\ a signal-to-noise of $5$ for an amplitude $\umax=0.1$ would mean an error bar of $0.1/5=0.02$ on that amplitude. We also show the correlation on the TTT bispectrum between the MAP of each mode with the \emph{constant feature} tested by Planck 2015 \cite[eq.\ (15)]{Ade:2015ava}, with phases corresponding to the cosine and sine cases, as well as the correlation with the ISW-lensing secondary bispectrum.
}
\end{table}

Direct searches for features have been performed in the bispectrum of both data releases of Planck, first using only the TTT bispectrum \cite{Ade:2013ydc} and later including polarisation and much higher oscillatory frequencies \cite{Ade:2015ava}. There, it was stated that oscillatory features that connected the aforementioned structure found in Planck's bispectrum achieved higher significance, but in neither of those cases a fit was found with a significance high enough to be called a detection; nevertheless the results from fits of oscillatory features were deemed ``interesting hints of non-Gaussianity''.

We can speculate whether our predictions are consistent with those hints. In particular, let us look at the linearly-oscillating templates tested by them, whose frequency very precisely satisfies $\omega\approx|\tau_0|$. None of the templates tested by Planck present the shape weighting and difference in phase between limits of our shapes, or their particular enveloping, that enhances the signal at high-$\ell$. So we focus on the simplest case of a \emph{constant feature} \cite[eq.\ (15)]{Ade:2015ava} $B(k_1,k_2,k_3)\propto\sin[\omega(k_1+k_2+k_3)+\phi]/(k_1k_2k_3)^2$, unenveloped and shape-agnostic. Its correlations with our modes \textit{100} and \textit{200} for a frequency $\omega\approx|\tau_0|$ and the sine and cosine phases are shown in table \ref{tab:corrs}. As we can see, the sine case is the most highly (anti) correlated one.\footnote{This is apparently at odds with the fact that most of our signal-to-noise comes from the squeezed limit, which would correspond to the cosine case; but not so surprising when one considers that, on top of the different enveloping, there is a strong difference in scaling towards the squeezed limit between our template (divergent at $\min_i(k_i)\rightarrow0$) and the constant feature (constant).}

Interestingly, the Planck collaboration do find a peak at $\omega\approx 100$, with a phase close to zero (especially in polarisation; the phase is not so small in temperature) and a negative amplitude with signal to noise of the expected order of magnitude ($\sim 0.5$ times the signal to noise of our templates). We find this coincidence tantalising, and look forward to testing our templates against Planck data directly in a joint search.

\section{Conclusions and discussion}
\label{sec:conclusion}

We have updated our ongoing search for features from transient reductions in the speed of sound of the inflaton with the new Planck 2015 polarised power spectrum data. We have proposed and explored a prior that exhausts the regime in which a feature coming from a Gaussian reduction in the speed of sound of the inflaton would be clearly distinguishable from the baseline cosmology. Since the prior is exhaustive and Planck's temperature power spectrum is cosmic-variance-limited for almost all the range that is relevant for inflationary features, we can consider these results definitive for the Gaussian ansatz, at least until higher signal-to-noise polarisation data is available for multipoles in the range $\ell=500$--$1500$.

We have found some modes that, though not statistically significant using power spectrum data only, have a very well constrained oscillation frequency and a rather well-defined amplitude, whereas their sharpness, in terms of $\smax$, is not constrained by the data but by the prior, which comes from the theoretical self-consistency. The predicted correlated bispectra of two of these modes show traits similar to those described in Planck's TTT bispectrum; in addition, Planck's search for linearly-oscillatory features picks up the frequency, sign and approximate phase of one of them.

This apparent similarity, though not at all conclusive, motivates us to repeat the present search in the future, including Planck's temperature and polarisation bispectra, and using the prior described in section \ref{sec:prior}. Such a search should also expand to regions of higher $|\tau_0|$ (higher oscillatory frequency) where nothing was found in the power spectrum -- the amplitude of the bispectrum features, contrary to that of the power spectrum's, is proportional to the oscillatory frequency due to the derivatives in eq.\ \eqref{eq:bispSRFT} \cite{Mooij:2015cxa}, and this significantly enhances the signal-to-noise of highly oscillatory features \cite{Munchmeyer:2014cca}. If the features corresponding to these modes are actually present in the data, combined searches in both the power spectrum and bispectrum are expected to greatly raise the significance of the fits \cite{Fergusson:2014hya,Meerburg:2015owa}, hopefully to detection-like levels.

If that combined search still fails to deliver enough significance, we will have to wait until larger tomographic data sets are available, such as 21 cm tomography \cite{Chen:2016zuu,Xu:2016kwz} or the next generation of Large Scale Structure surveys \cite{Chen:2016vvw,Ballardini:2016hpi}.

\begin{acknowledgments}
We thank Vicente Atal, Antony Lewis, Moritz M\"unchmeyer, Pablo Ortiz, Gonzalo Palma, Julien Peloton and Donough Regan for various helpful discussions. JT and BH thank the Institute Lorentz of the University of Leiden for its hospitality. JT acknowledges support from the European Research Council under the European Union's Seventh Framework Programme (FP/2007-2013) / ERC Grant Agreement No.\ [616170], and the Engineering and Physical Sciences Research Council [grant number EP/I036575/1]. BH is partially supported by the Chinese National Youth Thousand Talents Program and the Spanish Programa Beatriu de Pin\'os. AA acknowledges support by the Basque Government (IT-979-16) and the Spanish Ministry MINECO (FPA2015-64041-C2-1P). AA’s work was partially supported by the Simons Foundation, the Organization for Research in Matter (FOM), and the Netherlands Organization for Scientific Research (NWO/OCW).
\end{acknowledgments}

\appendix

\section{Bispectrum computation}\label{app:bispectrum}

\subsection{Review of expansion in total scale}

We attempt to apply the method proposed in \cite{Munchmeyer:2014cca}, based on an expansion in the total scale $k_t\defeq k_1+k_2+k_3$. There, one assumes that the primordial bispectrum can be written such that the shape function does depend on the total scale only, i.e.
\begin{equation}
  B_\curv(k_1, k_2, k_3) = \frac{(2\pi)^4\mathcal{A}_s^2}{(k_1k_2k_3)^2} S(k_t)
  \,.
\end{equation}
Then one expands the shape function in a Fourier series in an interval $[k_{t,\mathrm{min}}, k_{t,\mathrm{max}}]$ in whose extremes the shape function is zero, up to a sufficient order $n_\mathrm{max}$:
\begin{equation}
 S(k_t) = \sum_{n=0}^{n_\mathrm{max}} \left[
          \alpha_n c_n(k_t) + \beta_n  s_n(k_t)
        \right]\,,
\end{equation}
where we have abbreviated
\begin{equation}
  c_n(k) \defeq \cos\left(2\pi\,n\frac{k}{k_{t,\mathrm{max}}-k_{t,\mathrm{min}}}\right)
  \,,\qquad
  s_n(k) \defeq \sin\left(2\pi\,n\frac{k}{k_{t,\mathrm{max}}-k_{t,\mathrm{min}}}\right)
  \,.
\end{equation}
The coefficients of the Fourier series are
\begin{equation} 
  \alpha_n = \frac{2}{k_{t,\mathrm{max}}-k_{t,\mathrm{min}}}
             \int_{k_{t,\mathrm{min}}}^{k_{t,\mathrm{max}}}
             \diff k_t\,S(k_t) c_n(k_t)
  \,,\qquad
  \left(\alpha_n \rightarrow \beta_n\,,\;c_n(k_t) \rightarrow s_n(k_t)\right) 
  \,.
\end{equation}

A crucial advantage of this method is that the sine and cosine in the total scale are separable:
\begin{align}\label{eq:septrig}
  c_n (k_t)&= \phantom{-}c_n(k_1)c_n(k_2)c_n(k_3) - \left[s_n(k_1)s_n(k_2)c_n(k_3)+\cperms\right]\\
  s_n (k_t)&= -s_n(k_1)s_n(k_2)s_n(k_3) + 
  \left[c_n(k_1)c_n(k_2)s_n(k_3)+\cperms\right]
  \,,
\end{align}
where $\cperms$ means the 2 remaining cyclic permutations of the $k_i$.

Now, remember that the primordial bispectrum gets projected to the reduced CMB bispectrum as
\begin{equation}
  b_{\ell_1\ell_2\ell_3} = \left(\frac{2}{\pi}\right)^3 \int \diff r\, r^2
               \int \diff k_1\, \diff k_2\, \diff k_3\, (k_1k_2k_3)^2
               B_\curv(k_1, k_2, k_3) \prod_{i=1}^3 \Delta_{l_i}(k_i) j_{l_i}(k_ir)
  \,.
\end{equation}
Defining
\begin{equation}
  \mathcal{C}_{\ell n} \defeq \frac{2}{\pi}
  \int \diff k\, j_\ell(k r)\, \Delta_{\ell}(k)\, c_n(k)
  \,,\qquad
  \left(\mathcal{C}_{\ell n} \rightarrow \mathcal{S}_{\ell n}\,,\;c_n(k) \rightarrow s_n(k)\right) 
  \,.
\end{equation}
and, equivalently
\begin{align}
  \mathcal{C}_{\ell_1\ell_2\ell_3,n} &\defeq (2\pi)^4 \int \diff r\, r^2 \left[
                     \phantom{-}\mathcal{C}_{\ell_1 n}\mathcal{C}_{\ell_2 n}\mathcal{C}_{\ell_3 n} -
                     \left(\mathcal{S}_{\ell_1 n}\mathcal{S}_{\ell_2 n}\mathcal{C}_{\ell_3 n}+\cperms\right)\right]\\
  \mathcal{S}_{\ell_1\ell_2\ell_3,n} &\defeq (2\pi)^4 \int \diff r\, r^2 \left[
                              - \mathcal{S}_{\ell_1 n}\mathcal{S}_{\ell_2 n}\mathcal{S}_{\ell_3 n} +
                     \left(\mathcal{C}_{\ell_1 n}\mathcal{C}_{\ell_2 n}\mathcal{S}_{\ell_3 n}+\cperms\right)\right]
  \,,
\end{align}
where $\cperms$ means the 2 remaining cyclic permutations of the $\ell_i$. The final reduced bispectrum is
\begin{equation}
  b_{\ell_1\ell_2\ell_3} = \mathcal{A}_s^2 \sum_{n=0}^{n_\mathrm{max}}
  \left( \alpha_n \mathcal{C}_{\ell_1\ell_2\ell_3,n} + \beta_n \mathcal{S}_{\ell_1\ell_2\ell_3,n}\right)
  \,.
\end{equation}
The reduced bispectrum is thus \emph{separable}, but there is an additional advantage: whatever model parameters the primordial shape depends upon are now contained in the Fourier coefficients $\alpha_n$ and $\beta_n$ (and, indirectly, in the choice of $n_\mathrm{max}$ and the interval $[k_{t,\mathrm{min}}, k_{t,\mathrm{max}}]$). Thus, if we want to compute the CMB bispectrum for different values of the primordial model parameters, while keeping the background cosmology unchanged, we only need to re-calculate the Fourier coefficients, and we can re-use already pre-computed and stored, projected Fourier modes $\mathcal{C}_{\ell_1\ell_2\ell_3,n}$ and $\mathcal{S}_{\ell_1\ell_2\ell_3,n}$

\subsection{Extension and applicability to our bispectrum template}

Let us now write a slightly more complicated template:
\begin{equation}
  B_\curv(k_1, k_2, k_3) = \frac{(2\pi)^4\mathcal{A}_s^2}{(k_1k_2k_3)^2}
  \left[f(k_1) g(k_2) h(k_3) +\perms\right] S(k_t)
  \,,
\end{equation}
where $\perms$ here runs over the possible combinations of the three functions and the three momenta. For symmetry reasons, this is the way a separable factor would take; e.g. the simplest case would be $k_1+k_2+k_3$, which corresponds to $f=k/2,\,g=h=1$, and $k_1k_2k_3$ would correspond to $f=g=h=6^{{-1}/{3}}k$ (these decompositions are not unique).

Now let's promote the $\mathcal{C}_{\ell n}$ and $\mathcal{S}_{\ell n}$ to operators over functions of a single $k_i$:
\begin{equation}
  \label{eq:intsingle}
  \mathcal{C}_{\ell n}[f] \defeq \frac{2}{\pi}
  \int \diff k\, j_\ell(k r)\, \Delta_{\ell}(k)\, c_n(k)\, f(k)
  \,,\qquad
  \left(\mathcal{C}_{\ell n}[f] \rightarrow \mathcal{S}_{\ell n}[f]\,,\;c_n(k) \rightarrow s_n(k)\right) 
  \,.
\end{equation}
and, equivalently
\begin{align}
  \label{eq:intcomb}
  \mathcal{C}_{\ell_1\ell_2\ell_3,n}[f,g,h] &\defeq (2\pi)^4 \int \diff r\, r^2 \sum_{\left(f,g,h\right)}
    \left[\phantom{-}\mathcal{C}_{\ell_1 n}[f]\mathcal{C}_{\ell_2 n}[g]\mathcal{C}_{\ell_3 n}[h] -
          \left(\mathcal{S}_{\ell_1 n}[f]\mathcal{S}_{\ell_2 n}[g]\mathcal{C}_{\ell_3 n}[h]+\cperms\right)\right]\\
  \mathcal{S}_{\ell_1\ell_2\ell_3,n}[f,g,h] &\defeq (2\pi)^4 \int \diff r\, r^2 \sum_{\left(f,g,h\right)}
    \left[         - \mathcal{S}_{\ell_1 n}[f]\mathcal{S}_{\ell_2 n}[g]\mathcal{S}_{\ell_3 n}[h] +
          \left(\mathcal{C}_{\ell_1 n}[f]\mathcal{C}_{\ell_2 n}[g]\mathcal{S}_{\ell_3 n}[h]+\cperms\right)\right]
  \,,
\end{align}
where the sum runs over all 6 permutations of the three functions $f$, $g$, $h$. In this case, the total reduced bispectrum would be
\begin{equation}
  b_{\ell_1\ell_2\ell_3} = \mathcal{A}_s^2 \sum_{n=0}^{n_\mathrm{max}}
  \left( \alpha_n \mathcal{C}_{\ell_1\ell_2\ell_3,n}[f,g,h] + \beta_n \mathcal{S}_{\ell_1\ell_2\ell_3,n}[f,g,h]\right)
  \,.
\end{equation}
If we have more terms with said structure, we can recover the full bispectrum by just summing them over.

At this point, one may wonder how much complication we have introduced with respect to the method presented in \cite{Munchmeyer:2014cca}. To see that, let's detail the expected computational sequence if we want to obtain the full bispectrum:
\begin{enumerate}
\item Compute the $\mathcal{C}_{\ell n}$ and $\mathcal{S}_{\ell n}$ for each $\ell$ and $n$ we are interested in. In this extension, this must be done at worst three times per term, for three different $f$, $g$, $h$ per term. That is at worst $3 n_\text{terms}$.
\item Further integrate on $r$ the necessary combinations of the $\mathcal{C}_{\ell n}$ and $\mathcal{S}_{\ell n}$ to get the $\mathcal{C}_{\ell_1\ell_2\ell_3,n}$ and $\mathcal{S}_{\ell_1\ell_2\ell_3,n}$. In this case, this must be done once per term, times the 3 cyclic combinations of the functions, so $3 n_\text{terms}$ again.
\item Decompose the shape functions $S(k_t)$ on Fourier modes and sum those over the $\mathcal{C}_{\ell_1\ell_2\ell_3,n}$ and $\mathcal{S}_{\ell_1\ell_2\ell_3,n}$. In the extension, this must be done once per term: $n_\text{terms}$ slower.
\end{enumerate}
Thus, given that the two first steps are the ones that take longest by far and dominate the computation time, the extension is $3 n_\text{terms}$ as slow (or a smaller number of times $n_\text{terms}$ if two or all three of $f$, $g$ and $h$ are the same, or if they are the same within a term or between terms).

However, if we are not interested in varying the background cosmology, and if the parameters of the primordial model enter through $S(k_t)$ only or as some external intensity factor $\fnl$, but not through the functions $f$, $g$ and $h$, steps 1 and 2 can be pre-computed and stored. In that case, if one leaves the background cosmology unchanged, only step 3 is carried out and this method is only $n_\text{terms}$ slower. Notice that the choice of the maximum order $n_\mathrm{max}$ and the interval on which the Fourier decomposition is carried out depend implicitly on the model parameters; e.g. a higher frequency oscillation of the primordial shape requires a higher $n_\mathrm{max}$, which may not have been pre-computed yet.

In any case, remarkably, the computational costs grow only linearly with the number of terms, allowing us to compute more complicated non-separable bispectra that account for a richer set of physical scenarios.

\subsection{Notes on precision}

Every step in the computation described above, 1 to 3, has its own considerations regarding to precision: first the computation of the integrals of the $\mathcal{C}_{\ell n}$ and $\mathcal{S}_{\ell n}$, then the computation of $\mathcal{C}_{\ell_1\ell_2\ell_3,n}$ and $\mathcal{S}_{\ell_1\ell_2\ell_3,n}$, and finally, the Fourier decomposition of the shape functions.

\subsubsection{Computation of $\mathcal{C}_{\ell n}$ and $\mathcal{S}_{\ell n}$}

We want to compute the integrals in eq.\ \eqref{eq:intsingle} with enough precision.
They have four elements: three oscillatory functions (a transfer function, a spherical Bessel function and a sine or cosine) and a coefficient function of a single $k$.

To begin with, let us assume that the coefficient functions are very smooth compared to the rest of the oscillatory factors, so we can care about the sampling of the oscillations only. This is a reasonable hypothesis at least in the case that we are considering: they are monomials of a low degree.

For the oscillatory functions, we will parametrise the integration precision though an \emph{adaptive} parameter $N_I$, meaning the number of samples per oscillation of the fastest oscillator for each combination of $\ell$ and $n$. In this paper, we use $N_I = 20$, which should be enough for most purposes. The integrals are performed using a Simpson integrator.

\paragraph{Spherical Bessel functions:} They behave in the asymptotic limit as
\begin{equation}
\lim_{x\rightarrow 0} j_\ell(x) \sim x^\ell
\qquad\qquad
\lim_{x\rightarrow \infty} j_\ell(x) \sim \frac{1}{x} \cos\left(x-(\ell+1)\frac{\pi}{2}\right)
\,.\end{equation}
Therefore, to be sure to have $N_I$ samples per oscillation, it is enough to have a $\delta x={2\pi}/{N_I}$. In our particular case, the function is evaluated at $x\defeq rk$, which means that the wavelength along $k$ is $2\pi/r$. The distance along the line of sight, $r$ is evaluated later over a small interval around recombination. The worst-case scenario, the shortest wavelength, corresponds to the maximum $r$ of that interval. Thus, the desired sampling density on $k$ will be
\begin{equation}
\delta k = \frac{2\pi}{N_I r_\mathrm{max}}\,,
\end{equation}
where $r_\mathrm{max}$ is the maximum sampling value for the distance to recombination. For a reasonable value of $r_\mathrm{max} \simeq 1.5\cdot 10^4\,\mathrm{Mpc}$, this means $\delta k \simeq 2\cdot 10^{-5}\ \mathrm{Mpc^{-1}}$.

\paragraph{Transfer functions:} They are roughly proportional to $j_\ell(r_\mathrm{re}k)$, so the sampling strategy is the same as the last one, since $r_\mathrm{max}$ is sampled very close to recombination, $r_\mathrm{max}\simeq r_\mathrm{re}$.

\paragraph{Fourier series basis:} Since the wavelength of the basis function of order $n$ is $(k_{t,\mathrm{max}}-k_{t,\mathrm{min}})/n$, the necessary sampling here is
\begin{equation}
\delta k = \frac{k_{t,\mathrm{max}}-k_{t,\mathrm{min}}}{N_I n}
\,.
\end{equation}
Notice that for an interval of order 0 length we would need to go to an order $n\sim 10^4$ in the Fourier expansion for the basis functions to oscillate faster than the transfer and spherical Bessel functions. Thus, it will be the last ones that in most cases will impose the sampling density.

\subsubsection{Computation of $\mathcal{C}_{\ell_1\ell_2\ell_3,n}$ and $\mathcal{S}_{\ell_1\ell_2\ell_3,n}$}

The next step is to compute the integral along the line of sight from the recombination epoch to the present day in eq. \eqref{eq:intcomb}. Roughly speaking, the CMB temperature anisotropy is mainly produced by the inhomogeneities on the last-scattering surface, after photon-electron decoupling, so that the CMB photons are almost free to propagate until they reach us today. Mathematically, this means that each multipole of the CMB radiation transfer functions behaves like a Dirac delta function centred at $\ell \sim r_\mathrm{rec} k$. Hence, it is enough to sample a thin interval around $r_\mathrm{rec}\simeq 1.4\cdot10^4\,\mathrm{Mpc}$~\cite{Smith:2006ud}. In this work, we sample $200$ points linearly spaced in the interval $[1.3304, 1.5284]\cdot10^4\,\mathrm{Mpc}$, and integrate using a Simpson quadrature.

\subsubsection{Fourier decomposition: limits, order and sampling in $\ell$}

\paragraph{Limits of the Fourier decomposition:} In the choice of the limits of the interval of the Fourier decomposition, $[k_{t,\mathrm{min}},\,k_{t,\mathrm{max}}]$, we must take into account two things. First, it is preferable that the shape functions are zero in both extremes of the interval; otherwise we can expect Gibbs overshoots and ringing in the extremes of the interval, and in order to suppress them we would need to go to higher order of the Fourier series, adding computational costs. Second, since the Fourier reconstruction is periodic outside the interval, if the interval is smaller than the sampling interval of the transfer functions, we may find copies of the shapes at higher $k$'s. Given that the maximum $k_t$ reached is three times the maximum $k_i$ sampled in the transfer functions, that have a sampling interval of roughly $k_i\in[10^{-6},\,0.35]\,\mathrm{Mpc}^{-1}$, a good choice is to take the decomposition interval as $k_t\in[0,\,1]\,\mathrm{Mpc}^{-1}$. As long as the Gibbs artefacts happen mostly at the end of the interval, they would be hidden by the low value of the transfer functions there.

\paragraph{Maximum order of the Fourier series:} The order of the Fourier series must be high enough to correctly represent the shape functions, which is completely model dependent. This is best checked by directly comparing the original bispectrum shape with the reconstruction from the Fourier decomposition.

\paragraph{Sampling in $\ell$ space:} Approximately, $\ell \simeq r_\mathrm{re} k$; thus, if a sampling density $\delta k$ accurately represents the primordial bispectrum, a corresponding $\delta \ell = r_\mathrm{re} \delta k$ should provide a good sample of the bispectrum. For computational feasibility, we may reduce the sampling density in $\ell$ and construct a sufficiently accurate spline approximation, in order to calculate e.g. the signal-to-noise.

\subsection{Application to our case}

Let us repeat the equation of the primordial bispectrum:
\begin{equation}
  \label{eq:bispSRFTparams}
  B_\curv(k_1, k_2, k_3) = \umax\,\frac{(2\pi)^4\mathcal{A}_s^2}{(k_1k_2k_3)^2}
  \sum_{i=0}^2 c_i(k_1, k_2, k_3) \left(\frac{\diff}{\diff \sfrac{k_t}{2}}\right)^i \dppt(\sfrac{k_t}{2};\,\smax,\tau_0)
  \,,
\end{equation}
where $k_t \defeq k_1+k_2+k_3$. The $c_i$ coefficients are given in equation \eqref{eq:bispeccoeffs}.

We have made explicit the dependence on the parameters of the reduction in $\cs$,  $(\umax, \smax, \tau_0)$, to highlight an important property of this expression:
\begin{itemize}
\item $\umax$ enters only as an overall factor (the combination $\umax^{-1}\dpp$ does not depend on it).
\item $\smax$ and $\tau_0$ enter only through $\umax^{-1}\dpp$.
\end{itemize}
The rest of the factors do not depend on the particular choice for the reduction. Thus, we can pre-compute all the integrals $\mathcal{C}_{\ell_1\ell_2\ell_3,n}$ and $\mathcal{S}_{\ell_1\ell_2\ell_3,n}$ for all function combinations, which amounts for most of the computing time for the full CMB bispectrum calculation. Later, for a particular choice of parameter values for the reduction, we can very quickly compute the Fourier decomposition of the total-momentum-dependent part, and sum the terms using the pre-computed integrals.

We also need to care about two model-dependent aspects of the precision: the maximum order of the Fourier decomposition $n_\mathrm{max}$ and the necessary sampling in $\ell$ (so we can interpolate for a faster computation of S/N). The determinant quantity in both cases is the maximum oscillatory frequency of the feature, and thus the maximum $|\tau_0|$ that we are interested in.

For $n_\mathrm{max}$, notice that the \emph{characteristic} order of the feature, i.e.\ that whose Fourier frequency is equals the oscillatory frequency of the feature, is $|\tau_0| (k_{t,\mathrm{max}}-k_{t,\mathrm{min}}) (2\pi)^{-1}$. Decomposing the shape functions up to twice that frequency should provide us with a good reconstruction. For the sake of safety, we may increase that order by a small security factor:
\begin{equation}
n_\mathrm{max} = {|\tau_0| (k_{t,\mathrm{max}}-k_{t,\mathrm{min}})}\frac{1}{\pi}(1+\epsilon)\,.
\end{equation}
For the maximum $|\tau_0|$ we are interested in in this paper, $1600$, an interval of $1\,\mathrm{Mpc}^{-1}$, as stated in the previous section, and a security factor of $1+\epsilon=1.5$, the maximum order that we need to pre-compute is $n_\mathrm{max}=765$ ($400$ is enough for the high $\beta$ regime).

Regarding the sample in the $\ell$-space, notice first that, approximately, $\ell \simeq r_\mathrm{re} k$. In the $\ell$-space, we aim at correctly sampling the feature oscillation. With the relationship stated above, the wavelength in $\ell$ of the feature is, approximately, $r_\mathrm{re}\,2\pi/|\tau_0|$. This means that if we wanted a reasonable amount of 20 samples per feature oscillation, to get a good interpolation, we would need a sampling $\Delta \ell \simeq 3$.

Here, we may notice that we can take advantage of the two different regimes, $\smax\gg\umax$ and $\smax\approx\umax$, discussed in section \ref{sec:theoryreview}: the features of highest $\beta$ ($\smax\gg\umax$) have also a lower $|\tau_0|$, at most around $820$, which means that $\Delta \ell \simeq 5$ is enough sampling. On the other hand, the features at lowest $\beta$ ($\smax\approx\umax$), though they present a higher $|\tau_0|$ and thus need a sampling of $\Delta \ell \simeq 3$ or smaller, are dead by $k_t=0.03\,\mathrm{Mpc}^{-1}$, which corresponds roughly to $\ell_t = \ell_1+\ell_2+\ell_3\simeq 400$. Thus we can sample more finely up to that $\ell_t$, and more coarsely after it, allowing for higher computational efficiency.

\subsection{Fisher matrix elements, signal-to-noise and correlations}
\label{sec:fisher}

We will use the spectra predicted/obtained to compute signal-to-noise and shape correlations through the Fisher matrix \cite{Smith:2006ud} between two temperature bispectra $i$ and $j$, assuming homogeneous noise:
\begin{equation}
  \label{eq:fisher}
  F_{ij} \defeq \sum_{\ell_\mathrm{min}\le\ell_1\le\ell_2\le\ell_3\le\ell_\mathrm{max}} \frac{2}{\pi}
  \left(\ell_1+\frac{1}{2}\right)\left(\ell_2+\frac{1}{2}\right)\left(\ell_3+\frac{1}{2}\right)
  \left(\begin{array}{ccc}\ell_1&\ell_2&\ell_3\\0&0&0\end{array}\right)^2
  \frac{b_{\ell_1\ell_2\ell_3}^{(i)}b_{\ell_1\ell_2\ell_3}^{(j)}}{\sigma^2_{\ell_1\ell_2\ell_3}}
\,,
\end{equation}
where $\ell_\mathrm{min}$ and $\ell_\mathrm{max}$ are the minimum and maximum values allowed for individual multipoles, here respectively $2$ and $2000$, and $\sigma^2_{\ell_1\ell_2\ell_3}$ is the approximate cosmic variance for a small bispectrum \cite{Luo:1993xx,Heavens:1998jb,Spergel:1999xn,Gangui:1999vg}
\begin{equation}
  \label{eq:bispeccosmicvariance}
  \sigma^2_{\ell_1\ell_2\ell_3} \approx C_{\ell_1}C_{\ell_2}C_{\ell_3}\Delta_{\ell_1\ell_2\ell_3}/f_\mathrm{sky}
\,,
\end{equation}
where $C_{\ell_i}$ is the \emph{observed} spectrum $C_{\ell,\mathrm{obs}} = C_{\ell,\mathrm{theo}}+N_\ell b_\ell^{-2}$, where we have assumed an effective Gaussian beam with $\theta_\mathrm{FWHM}=7.25\,\mathrm{arcmin}$ and white noise with standard deviation $\sigma_N = 33\,\mathrm{\mu K}\,\mathrm{arcmin}$, and the map mask leaves a sky fraction $f_\mathrm{sky}=0.76$. The geometrical factor $\Delta_{\ell_1\ell_2\ell_3}$ enforces the triangular condition on the three $\ell$'s and evaluates to $1,2,6$ respectively for the cases of all $\ell$'s being different, two being equal and all being equal. For $i$ and $j$ being two different bispectra, we assume the same background cosmology, and only a different inflationary model. We compute the necessary Wigner 3-$j$ symbol using the WIGXJPF algorithm \cite{Johansson:2015cca}.

From this Fisher matrix, one can derive the signal-to-noise and the correlation between two bispectra $i$ and $j$ as \cite{Komatsu:2001rj}
\begin{equation}
  \label{eq:bispecstn}
  \left(\stn\right)_i = \sqrt{F_{i,i}}
  \qquad\text{and}\qquad
  \mathcal{C}_{i,j} = \frac{F_{i,j}}{\sqrt{F_{i,i}F_{j,j}}}
\,.
\end{equation}
Notice that the correlation between two bispectra is independent of the amplitude of either.

We compute the Fisher matrix elements summing by slices of constant $\ell_t\defeq\ell_1+\ell_2+\ell_3$, with $\ell_i\in[2,2000]$, and interpolating the values we have not sampled in our bispectrum computation. We sample as many slices as we can within the target error for the Fisher matrix elements ($5$--$10\%$).

In principle, we could have taken advantage of the actual separability of the modal-expanded shape to pre-compute some of the steps of the Fisher matrix computation, and even maybe to create a KSW estimator, as is done in \cite{Munchmeyer:2014cca}. However, we choose not to do so in the present work: the amount of pre-computation needed is high, due to the sizeable number of different combinations of $c_i$ coefficients, and for now we do not want to streamline the Fisher matrix element computations -- we are not scanning the parameter space of the feature, but just computing signal-to-noise and correlations for particular parameter combinations.

\bibliographystyle{apsrev}
\bibliography{feature16} 

\end{document}